\documentclass[useAMS,usenatbib]{mn2e}
\usepackage{graphicx}
\usepackage{amsmath}
\usepackage{xcolor}
\usepackage{color}
\usepackage{ amssymb }
\usepackage{gensymb}

\newcommand{\be}{\begin{equation}}
\newcommand{\ee}{\end{equation}}

\newcommand{\ifm}[1]{\relax\ifmmode#1\else$\mathsurround=0pt #1$\fi}
\newcommand{\kms}{\ifmmode\,{\rm km}\,{\rm s}^{-1}\else km$\,$s$^{-1}$\fi}

\newcommand{\ltsima}{$\; \buildrel < \over \sim \;$}
\newcommand{\lsim}{\lower.5ex\hbox{\ltsima}}
\newcommand{\gtsima}{$\; \buildrel > \over \sim \;$}
\newcommand{\gsim}{\lower.5ex\hbox{\gtsima}}

\definecolor{green}{rgb}{0,0.5,0}
\definecolor{grey}{rgb}{0.4,0.5,0.7}

\def\M11{M_{11}}
\def\V100{V_{100}}
\def\R1{R_{Mpc}}
\def\T6{T_6}

\begin{document}

\title[Bursting and quenching in satellite galaxies]{Bursting and quenching in satellite galaxies}

\pagerange{\pageref{firstpage}--\pageref{lastpage}} \pubyear{2016}

\author[Koutsouridou and Cattaneo]{I.~Koutsouridou$^{1}$, A.~Cattaneo$^{1,2}$
\\
\\
$^1$Observatoire de Paris, GEPI, PSL University, 61 avenue de l’Observatoire, 75014 Paris, France \\
$^2$Institut d'Astrophysique de Paris, CNRS, 98bis Boulevard Arago, 75014 Paris, France\\
}

\maketitle

\label{firstpage}


\begin{abstract}

The difference in stellar metallicity between red and blue galaxies with the same mass constrains the timescale over which red galaxies ceased to form stars. Here we investigate this constraint with the {\sc GalICS~2.0} semi-analytic model of galaxy formation.
The advantage of this approach is that the time of pericentric passages for satellite galaxies and the mass-loading factor for galactic winds are not free parameters of the chemical evolution model.
The former is determined by the N-body simulation used to construct the merger trees, the latter by the requirement that  {\sc GalICS~2.0} should reproduce the stellar mass function of galaxies.
When we compare our theoretical predictions with observations, we find that {\sc GalICS~2.0} can reproduce the observed metallicity difference only if quenching is preceded by a burst of star formation,
which contributes to the chemical enrichment of the stellar population. Physically, this burst can be explained as tidally-induced star formation or as an effect of ram pressure,
which not only strips gas from galaxies but also compresses it, accelerating its conversion into stars.

\end{abstract} 


 
\section{Introduction}

The observation that red galaxies have higher metallicities than blue galaxies constrains the timescale over which the former ceased to form stars (\citealp{peng_etal15}; \citealp{trussler_etal18}; hereafter P15 and T18, respectively). If star formation were shut down instantaneously, the stellar metallicities of red galaxies would freeze to their  values at the time of quenching. In a more realistic scenario where the accretion of gas is shut down but galaxies keep forming stars until they run out of gas, the metallicity of the gas increases with time due to metal enrichment by supernovae (SNe) and the absence of any dilution; hence, stars with higher and higher metallicities are formed while the star formation rate (SFR) decays exponentially.
Observations strongly favour the second picture, especially at $M_*<10^{11}\,M_\odot$ (P15). 

The average gas and stellar metallicity, gas fraction and SFR for a blue galaxy of a given stellar mass are known from observations \citep{maiolino_etal08,Boselli2014,Popping2014,tacconi_etal18}.
T18 used these empirical initial conditions and a closed-box chemical-evolution model \citep{tinsley80} to compute how the stellar metallicity evolves after a galaxy ceases to accrete and starts moving to the red sequence. They found metallicities that match the observations after $\sim 2\,$Gyr without accretion, consistent with a delayed quenching scenario, in which  a slow exponential decline of the SFR is followed by abrupt quenching after $t_{\rm delay}\sim 2\,$Gyr.

Outflows are the main uncertainty in this calculation. If gas is blown out, there is less star formation and chemical enrichment  proceeds less rapidly. 
Therefore, more protracted star formation is necessary to produce the same final metallicity.
Let $\eta$ be the ratio of the outflow rate to the SFR. Mass-loading factors $\eta\gsim 1$ are incompatible with the large stellar metallicities of massive  galaxies. In dwarf galaxies, the situation is far less clear. Hence, the degeneracy between
$\eta$ and $t_{\rm delay}$ is more difficult to unravel.

The  real question, however, is what physical processes starved red galaxies of gas. Environmental effects such as tidal forces and ram pressure are the prime suspects. These effects can act two ways: if  they are weak, they disrupt the flow of gas onto galaxies; if they are  strong, they remove gas from galaxies. In the first case, they act as  a starvation mechanism and raise the metallicity; in the second case, they act as a quenching mechanism and freeze it. For a galaxy that falls into a cluster, tidal and ram-pressure stripping will be weaker at the entry time and stronger at the pericentre. Therefore, it is logical to associate the entry time with the time $t_{\rm s}$ at which gas accretion is shut down, and the pericentric passage with the time $t_{\rm q}$ at which star formation is quenched, so that $t_{\rm delay}=t_{\rm q}-t_{\rm s}$.

Studies of how red and blue galaxies populate the phase-space of galaxy clusters  \citep{mahajan_etal11,wetzel_etal13,muzzin_etal14} and semi-empirical models based on abundance matching \citep{tollet_etal17} converge towards a similar delayed-quenching scenario.
After a galaxy falls into a cluster its SFR declines slowly. Only after a delay time $t_{\rm delay}\sim 2\,$Gyr that corresponds to the lapse of time  between entry and the first pericentric passage does the SFR go to zero in under $0.5\,$Gyr \citep{muzzin_etal14}. If star formation were quenched immediately as soon as a galaxy becomes a satellite, the masses of satellite galaxies would be underpredicted. Without any stripping, they would be overpredicted \citep{tollet_etal17}.

Our goal is to use the {\sc GalICS}~2.0 semianalytic model (SAM) of galaxy formation (\citealp{cattaneo_etal17}; hereafter C17) 
to connect the results of these complementary approaches within the unifying framework of a coherent theory of the formation and evolution of galaxies in a cosmological context.
The great advantage of this approach is that $\eta$ and $t_{\rm delay}$ cease to be free parameters: $t_{\rm delay}$ is determined by the N-body simulation used to construct the merger trees; $\eta$ has already been tuned to reproduce the stellar mass function of galaxies.

The structure of the article is as follows. In Section~2, we describe the {\sc GalICS}~2.0 SAM together with the few differences between C17 and the version used for this article.
In Section~3, we present our results for the conditional stellar mass function of satellite galaxies and the metallicities of red and blue galaxies and compare them with the observational measurements of \citet{Yang2012} and T18, respectively. Our results are based on models that include strangulation but do not include stripping (no gas or stars are removed from galaxies). In Section~4, we estimate the maximum effect that ram-pressure and tidal stripping could have on our results. In Sections~5 and 6, we discuss these results and summarize the conclusions of the article.
  
\section{The model}
\subsection{Merger trees}

The {\sc GalICS~2.0} SAM runs on dark-matter (DM) merger trees from cosmological N-body simulations. The simulation used for this article assumes the same cosmology \citep{planck_etal14} and the same initial conditions as the one used in C17. The (sub)halo finder and the algorithm used to construct merger trees are the same, too. The only differences considering DM are the N-body resolution, which has been increased from $512^3$ particles to $1024^3$ particles for a volume of $(100{\rm\,Mpc})^3$, and the treatment of subhaloes when they cease to be resolved by the halo finder.

C17 assumed that whenever a subhalo is no longer detected by the halo finder, the associated satellite galaxy merges with the central galaxy of the host halo. This assumption artificially accelerates mergers even though the effect is less prominent at higher resolution. Even state-of-the-art N-body simulations suffer from significant overmerging, mainly because of inadequate force softening, which causes excessive mass loss. As long as a subhalo is followed with sufficient mass and force resolution a bound remnant often survives \citep{Kazantzidis2004, Diemand2007a, Diemand2007b, Bosch2018a,Bosch2018b}. Overmerging was not a problem for a study of the Tully-Fisher relation in isolated galaxies (C17) but it is for an article on satellite galaxies. Here we follow \citeauthor{tollet_etal17} (\citeyear{tollet_etal17}, T17 hereafter) 
 and use a formula derived from cosmological hydrodynamic simulations to circumvent the problem.

A satellite galaxy first enters the virial sphere of a group or a cluster and then merges with the central galaxies. Let $t_{\rm merge}$ be the time between these two events. \citet{jiang_etal08}'s formula gives $t_{\rm merge}$ as a function of $M_{\rm s}/M_{\rm vir}$ and orbital eccentricity ($M_{\rm s}$ and $M_{\rm vir}$ are the masses of the subhalo and the host halo, respectively).
Subhaloes that are no longer detected by the halo finder, survive as ghosts until the closest pericentric passage to the time when $t_{\rm merge}$ elapses. This assumption ensures that our average merging times are consistent with those measured in cosmological hydrodynamic simulations and that mergers always occur at the pericentre.

Pericentric passages are determined by treating ghost subhaloes as test particles subject to two forces: the gravitational attraction of the host halo and the dynamical friction force (T17). The gravitational attraction of the host halo is computed by assuming that its density distribution is described by an NFW model \citep{navarro_etal97}. The dynamical friction force is computed using \citet{chandrasekhar43}'s formula and depends on $M_{\rm s}$, which decreases with time due to tidal stripping  (see T17 for details).

T17 demonstrated that considering ghost subhaloes is equivalent to increasing the number of particles by at least an order of magnitude. They used abundance matching to assign stellar masses to the haloes and subhaloes identified in the simulations with $512^3$ and $1024^3$ particles. The
simulation with $512^3$ particles lacked satellites with stellar masses below $10^{10}\,M_\odot$ compared to the simulation with $1024^3$. However, once ghost haloes were included in the analysis of the simulation with $512^3$, its results agreed with those of the simulation with $1024^3$ particles down to a stellar mass of $10^9\,M_\odot$. By including ghost subhaloes in our analysis of the simulation with $1024^3$ particles, we are confident that our conditional stellar mass functions should be robust down to stellar masses of $\sim 10^8\,M_\odot$.

\subsection{Gas accretion}

Let $M_{\rm vir}$ be the mass of a halo and let $M_{\rm accr}$ be the total mass of all the baryons that have accreted onto the halo since the Big Bang. Whenever $M_{\rm vir}$ increases, $M_{\rm accr}$ increases so that $M_{\rm accr}/M_{\rm vir}$ is always equal to the universal baryon fraction in all but the lowest mass haloes, where the ratio $M_{\rm accr}/M_{\rm vir}$ is reduced by feedback from cosmic photoionisation (C17)\footnote{$M_{\rm accr}/M_{\rm vir}$ can be higher than the universal baryon if $M_{\rm vir}$ decreases because we do not remove baryons from haloes when that happens.}.

A fraction $f_{\rm hot}$ of the gas that accretes onto a halo is shock-heated. We assume $f_{\rm hot}=0$ for $M_{\rm vir}\le 10^{10.7}\,M_\odot$, $f_{\rm hot}=1$ for $M_{\rm vir}\ge 10^{12.7}\,M_\odot$, and that
$f_{\rm hot}$ grows linearly with ${\rm log}\,M_{\rm vir}$ at intermediate masses. In {\sc GalICS}~2.0, galaxies grow through cold accretion only, hence this gas never cools. The rest accretes onto the central galaxy on a timescale equal to the freefall time $t_{\rm ff}$, where $t_{\rm ff}$ is the time to reach the centre of the halo starting from the virial radius with speed equal to the freefall speed from infinity\footnote{C17 used the ratio of the virial radius to the virial velocity. Our new definition accelerates the accretion of the cold gas in the filaments by a factor of order unity.}.

In the standard version of {\sc GalICS~2.0}, cold gas already within the halo keeps accreting onto the galaxy, independently of whether the galaxy is central or satellite (model $a$). In this article, we also consider a model with strangulation, in which there is no gas accretion onto satellites (model $b$). In this case, the gas in the satellite's halo is transferred to the host at the time when the satellite galaxy and its host halo's central galaxy finally merge.

Bulges form through mergers (classical bulges) and disc-instabilities (pseudobulges). The structural properties of bulges are computed as in C17 but 
they are irrelevant for this work because they do not affect the star-formation rate (SFR, Section~2.3) and therefore have no effect on the metallicity evolution of galaxies. 

\subsection{Star formation and feedback}

Let $t_{\rm SF}$ be the star-formation timescale over which (atomic+molecular) gas is converted into stars\footnote{Many authors call star-formation timescale the inverse specific SFR and gas-depletion timescale the time
$t_{\rm SF}$ in Eq.~(\ref{tSF}). We do not follow this convention  because we call gas-depletion timescale the one defined in Eq.~(\ref{t_dep}).}, 
so that:
\begin{equation}
{\rm SFR} = \frac{M_{\rm gas}}{t_{\rm SF}}.
\label{tSF}
\end{equation}
For discs, we follow C17 and assume:
\begin{equation}
t_{\rm SF} = 25\,t_{\rm orb} = 25\times {2 \pi r_{\rm d}\over v_{\rm c}},
\label{tSFdisc}
\end{equation}
where $v_{\rm c}$ is the circular velocity and $r_{\rm d}$ is the disc's exponential scale-length computed assuming conservation of angular momentum (\citealp{mo_1998} and C17).
Eq.~(\ref{tSFdisc})  gives $t_{\rm SF}\simeq 3.5\,$Gyr on average (in  agreement with \citealp{Boselli2014}) but it also predicts a dependence on 
the angular momentum of the DM, which affects the value of $r_{\rm d}$.
To explore how this dependence  may affect our results, we also consider a variant of model $b$, in which all discs have a constant star formation timescale of $t_{\rm SF}=3.5\,$Gyr (model $c$). 
For merger-driven starbursts, all models considered in this article assume a constant SF timescale of $t_{\rm SF}=0.35\,$Gyr (consistent with \citealp{Bigiel2008}).
The effects of secular disc instabilities are far less clear as they can both increase and reduce the SF efficiency. We therefore ignore them altogether and assume that pseudobulges formed through 
disc instabilities have the same SF timescale as discs. 

A fraction $R$ of the gas that forms stars is returned to the interstellar medium (ISM) through stellar winds and SN explosions ($R$ corresponds to the mass fraction in short-lived massive stars).
In the limit that the lifetime of these stars can be considered negligible (instantaneous-recycling approximation), we can assume that both the rate of stellar mass loss and the outflow rate are directly proportional to the SFR,
with constants of proportionality $R$ and $\eta$, respectively.

With these assumptions, the evolution of the gas mass $M_{\rm gas}$ within a galactic component (e.g., the disc or the bulge) is governed by the equation:
\begin{equation}
\label{Mgas}
\dot{M}_{\rm gas} = \dot{M}_{\rm accr}-(1-R+\eta)\frac{M_{\rm gas}}{t_{\rm SF}},
\end{equation}
where $\dot{M}_{\rm accr}$ is the accretion rate onto the component (in {\sc GalICS~2.0}, discs are the only component that can accrete gas).
Eq.~(\ref{Mgas}) admits the simple analytic solution:
\begin{equation}
\label{Mgas_evol}
M_{\rm gas}(t) = \dot{M}_{\rm accr} \tau + [M_{\rm gas}(0)-\dot{M}_{\rm accr} \tau]e^{-{t\over\tau}},
\end{equation}
\citep{Cole2000, Lilly2013, Dekel2014, peng2014b}, where:
\begin{equation}
\label{t_dep}
\tau = {t_{\rm SF}\over 1-R+\eta}
\end{equation}
is the gas-depletion timescale (the reservoir of gas available for star formation is depleted not only by star formation but also by outflows).
Eq.~(\ref{Mgas_evol}) implies $M_{\rm gas}\rightarrow \dot{M}_{\rm accr} \tau$ for $t\gg\tau$.

By substituting Eq.~(\ref{Mgas_evol}) into Eq.~(\ref{tSF}), we find a similar equation for the evolution of the stellar mass:
\begin{equation} \label{Mstars_evol}
\begin{split}
M_*(t) & =  M_*(0) + (1-R)\int_0^t{\rm SFR}(t){\rm\,d}t \\
& =  M_*(0) + (1-R){\tau \over t_{\rm SF}}\cdot \\
\: \: \: \: \: \: \: \: \: \: \: \: \: \:\: \: \: \: \: \: \: \: & \cdot \Big[\dot{M}_{\rm accr}  t-(\dot{M}_{\rm accr} \tau -M_{\rm gas}(0))(1-e^{-{t\over\tau}} ) \Big].
\end{split}
\end{equation}

In {\sc GalICS~2.0}, the mass-loading factor $\eta$ is determined by the condition that our SAM should reproduce the stellar mass function of galaxies and its evolution with redshift.
C17 found the best-fit to observations \citep{yang_etal09,baldry_etal12,bernardi_etal13,moustakas_etal13,ilbert_etal13,muzzin_etal13,tomczak_etal14} for:
\begin{equation}
\eta=3.8\left(M_{\rm vir}\over 10^{11}\,M_{\odot}\right)^{-2},
\label{eta_galics}
\end{equation}
(models $a$, $b$ and $c$), where $M_{\rm vir}$ is the virial mass of the DM halo. We also consider a variant of model $b$ with $\eta=0$ in all satellite galaxies (model $d$).
Table~1 summarises the four models considered in this article.

{\sc GalICS~2.0} uses Eqs.~(\ref{Mgas_evol}) and~(\ref{Mstars_evol}) with the mass-loading factor in Eq.~(\ref{eta_galics}) 
to compute the evolution of the gas mass and the stellar mass in each galactic component 
from one timestep in the merger-tree to the next.

\subsection{Chemical enrichment}

\begin{table}
\begin{center}
\caption{Models considered in this article. Accretion and feedback refer to satellite galaxies only. Star formation refers to the total gas (atomic+molecular) mass of the disc component.}
\begin{tabular}{ c c c c}
\hline
\hline 
Model & Accretion & Star formation & Feedback \\
\hline
$a$&Standard               &Eq.~(\ref{tSFdisc})             &Eq.~(\ref{eta_galics})\\
$b$&No                         &Eq.~(\ref{tSFdisc})             &Eq.~(\ref{eta_galics})\\
$c$&No                         &$t_{\rm SF}=3.5{\rm\,Gyr}$&Eq.~(\ref{eta_galics})\\
$d$&No                         &Eq.~(\ref{tSFdisc})             &$\eta=0$\\
\hline
\hline
\end{tabular}
\end{center}
\label{model_parameters}
\end{table}

The mass of metals in the gas, $M_{\rm gas,Z}$, follows a similar equation to the one for $M_{\rm gas}$, with the difference that metals are not only accreted, used to form stars, returned to the ISM and ejected;
metals can also be produced through stellar nucleosynthesis. Let $\Delta M_{\rm Z}$ be the mass of metals produced and dispersed into the ISM by SNe and stellar winds when the stellar mass increases by $\Delta M_*$ due to star formation.
We define the metal yield $y=\Delta M_{\rm Z}/\Delta M_*$.  With this definition, the equation for the mass of metals is:
\begin{equation}
\label{MgasZ}
\dot{M}_{\rm gas,Z} = Z_{\rm IGM}\dot{M}_{\rm accr}-(1-R+\eta)\frac{M_{\rm gas,Z}}{t_{\rm SF}}+y(1-R)\frac{M_{\rm gas}}{t_{\rm SF}}
\end{equation}
\citep{tinsley80}, where $Z_{\rm IGM}$ is the metallicity of the accreted gas.

In the same way that \citet{Cole2000}, \citet{Lilly2013}, \citet{Dekel2014} and \citet{peng2014b} had found an analytic solution for Eq.~(\ref{Mgas}),
we have found that Eq.~(\ref{MgasZ}) admits the analytic solution:
\begin{equation}
M_{\rm gas,Z}(t) =(Z_{\rm IGM}+\tilde{y})\dot{M}_{\rm accr}\tau+
\label{MgasZ_evol}
\end{equation}
$$+[M_{\rm gas,Z}(0)-(Z_{\rm IGM}+\tilde{y})\dot{M}_{\rm accr}\tau]e^{-{t\over\tau}} +$$
$$-\tilde{y}\left[\dot{M}_{\rm accr}-{M_{\rm gas}(0)\over\tau}\right]te^{-{t\over\tau}},$$
where:
\begin{equation}
\tilde{y} ={1-R\over 1-R+\eta}y.
\end{equation}
Hence, $M_{\rm gas,Z}\rightarrow (Z_{\rm IGM}+\tilde{y})\dot{M}_{\rm accr}\tau$ for $t\gg\tau$.
We make our calculations simpler by assuming $Z_{\rm IGM}=0$, since the accreted gas is unlikely to have metallicity larger than one third the Solar value,
so that  $Z_{\rm IGM}\ll\tilde{y}$ for all galaxies in the range of masses considered in this article where $\eta$ is never much greater than unity.
With this assumption, we find:
\begin{equation}
Z_{\rm gas}={M_{\rm gas,Z}\over M_{\rm gas}}\sim {1-R\over 1-R+\eta}y
\label{equilibrium_y}
\end{equation}
(Eqs.~\ref{Mgas_evol},~\ref{MgasZ_evol}) as long as $\dot{M}_{\rm accr}$ is greater than zero\footnote{If gas accretion stops, $Z_{\rm gas}\rightarrow 1$ when $t \rightarrow \infty$.} and varies on timescales greater than $\tau$.

The mass of the metals locked into stars is:

\begin{equation} \label{MstarsZ_evol}
\begin{split}
 M_{*,Z}(t)& = M_{*,Z}(0) + (1-R)\int_0^tZ_{\rm gas}(t){\rm SFR}(t){\rm\,d}t \\
 & =M_{*,Z}(0) + (1-R)\frac{\tau}{t_{\rm SF}}\cdot \\
  \cdot \bigg{\{}( Z_{\rm IGM} &+  \tilde{y})\dot{M}_{\rm accr} t+\tilde{y}\left(\dot{M}_{\rm accr}-{M_{\rm gas}(0)\over\tau}\right)t e^{- {t\over\tau}}+ \\
 -\Big[(Z_{\rm IGM}+\tilde{y} & )\dot{M}_{\rm accr}\tau+\tilde{y}\left(\dot{M}_{\rm accr}-{M_{\rm gas}(0)\over\tau}\right)\tau-M_{\rm gas,Z}(0)\Big]\cdot \\
 & \: \: \: \: \: \: \: \: \: \: \: \: \: \: \: \: \: \: \: \: \: \: \: \: \: \: \: \: \: \: \: \: \:  \: \: \: \: \: \: \: \: \: \: \:  \: \:   \cdot\left(1-e^{-{t\over\tau}}\right)\bigg{\}},
\end{split}
\end{equation} 
so that the stellar metallicity, is simply found by computing $Z_*=M_{*,Z}/M_*$ (Eqs.~\ref{Mstars_evol} and \ref{MstarsZ_evol}).

$R$ and $y$ can be computed theoretically, by assuming a stellar initial mass function (IMF) and by using a stellar-evolution model
to compute returned fractions and metal yields as a function of stellar mass, or they can be determined observationally by fitting the $Z_*$ -- $M_*$ relation.

From a theoretical perspective, $R$ and $y$ are higher for IMFs with a higher fraction of massive stars. In this work we have adopted a \citet{Chabrier2003} IMF and  assumed it to be invariant with time and galaxy type/mass. T18 assumed a \citet{kroupa01} IMF, but their results do not change if one adopts the \citet{Chabrier2003} IMF.
\citet{Vincenzo2016}  assumed an  upper cut-off at $100\,M_\odot$ and found that the \citet{Chabrier2003} IMF gives returned fractions in the range of $0.436\lsim R\lsim 0.455$ for the stellar yields of \citet{Romano2010} and $0.403\lsim R\lsim 0.466$ for the yields of \citet{nomoto_etal13}. The corresponding yields are in the range of $0.049\lsim y\lsim 0.088$.
However, there is some evidence that the majority of massive stars above 25 -- 40$\,M_\odot$ may collapse to black holes that would swallow a significant part of their nucleosynthesis products \citep{smartt09,heger_etal0}\footnote{The transition may be smeared by metallicity, rotation and mass loss, and by the presence of an intermediate regime with weak SNe.},
in which case $R$ and $y$ will be lower than the values found by \citet{Vincenzo2016}.
The change in $R$ will be minor because stars above $25$ let alone $40\,M_\odot$ make up a small fraction of the total stellar mass but the effect on $y$ may be more significant because $25\,M_\odot$ is only three times the minimum mass for SN explosions \citep{smartt09}.
Eq.~(\ref{equilibrium_y}), however, shows that $y$ acts as a normalisation coefficient, which affects all galaxies in the same manner, and that limits the impact of the value of $y$ on our results.

{\sc GalICS~2.0} reproduces the $Z_*$ -- $M_*$ relation by \citet{Gallazzi2005} for $R=0.41$ and $y=0.034$ (Fig.~1, model $a$).  These parameters correspond to a maximum mass for SNe of $\sim 50\,M_\odot$, i.e. more massive stars eject no (enriched or unenriched) material into the ISM. We have calculated the aforementioned values by fitting the results of model $a$ to observational data including both star forming and passive galaxies. We expect the fitting to depend only weakly on the assumed model, since the median MZR is dominated by star forming galaxies, whose behaviour does not vary significantly between models. Nonetheless, a more detailed analysis should account for the radial metallicity gradient of galaxies, since the SDSS spectroscopic fibre
only samples the galaxy emission of the central regions of
galaxies at the redshift of interest.

\begin{figure}
\begin{center}
\includegraphics[width=1.\hsize]{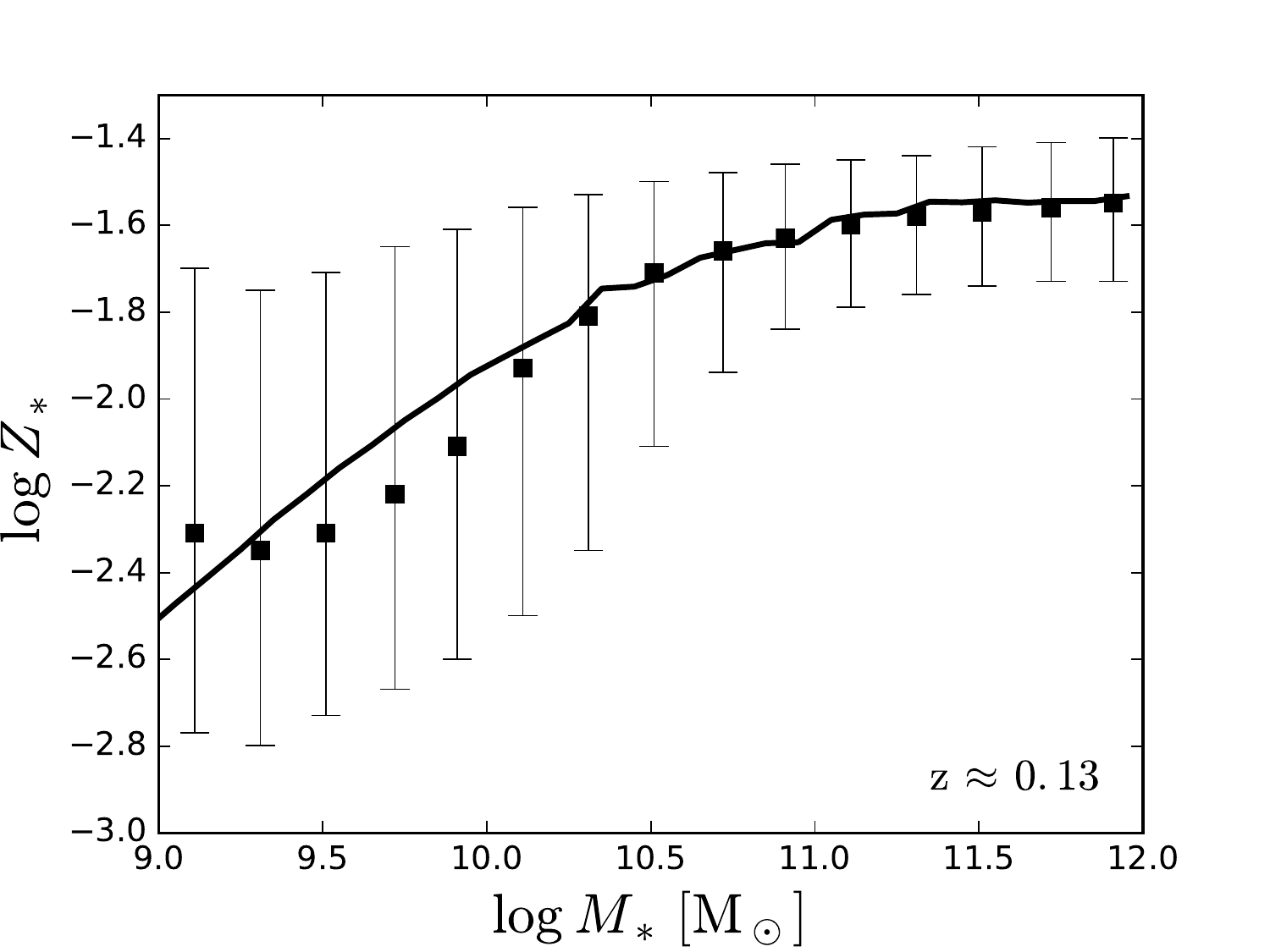} 
\end{center}
\caption{Median stellar metallicity as a function of stellar mass in {\sc GalICS~2.0} (model $a$ with $R=0.41$ and $y=0.034$; black curve) and in the observations by Gallazzi et al. (2005, data points; the error bars show the 16th and 84th percentiles).
The results of  {\sc GalICS~2.0} are shown at the same redshift as the data by Gallazzi et al. ($z\simeq 0.13$).}
\label{compMZR}
\end{figure}

\begin{figure*}
\begin{center}$
\begin{array}{cc} 
\includegraphics[width=0.5\hsize]{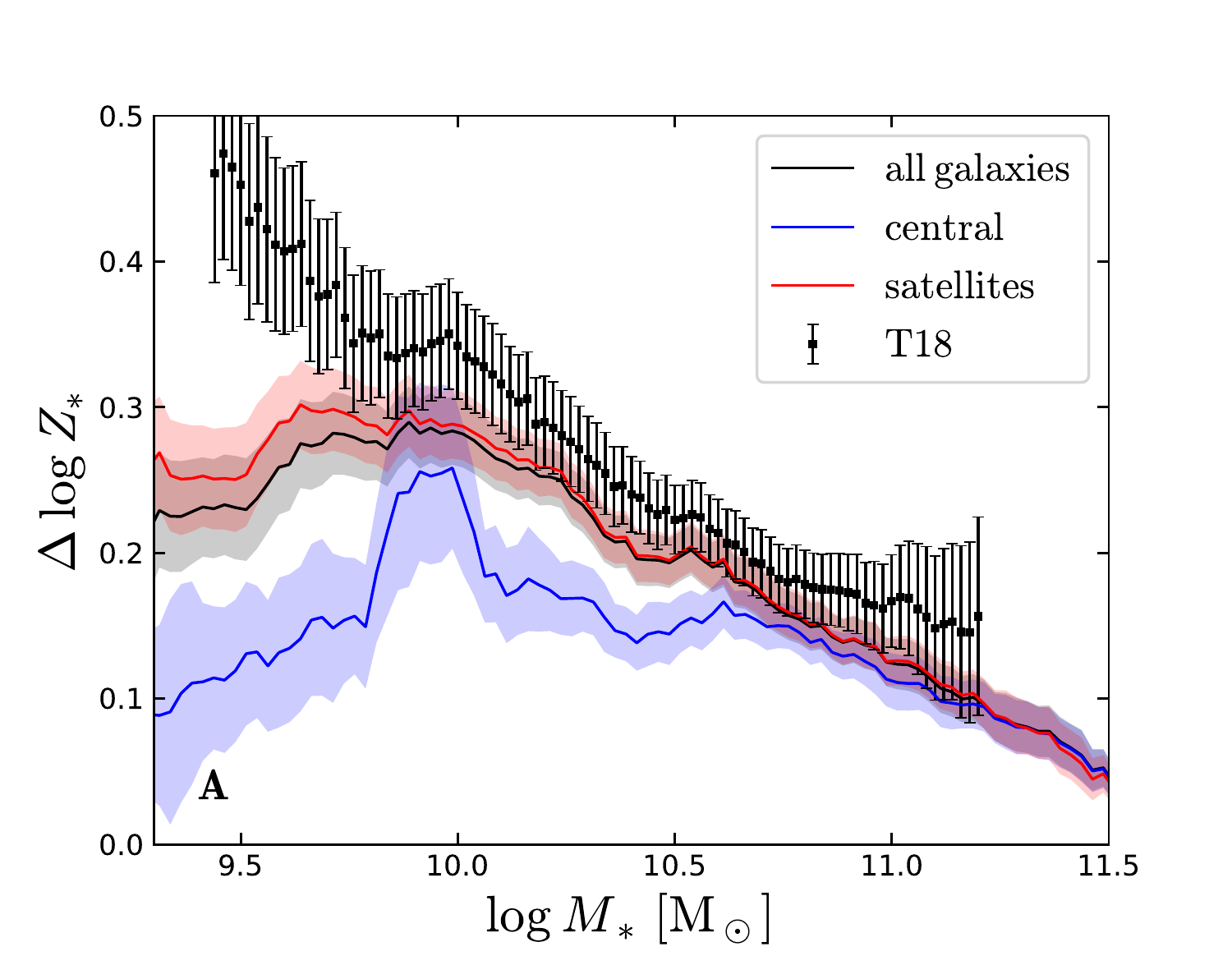} &
\includegraphics[width=0.5\hsize]{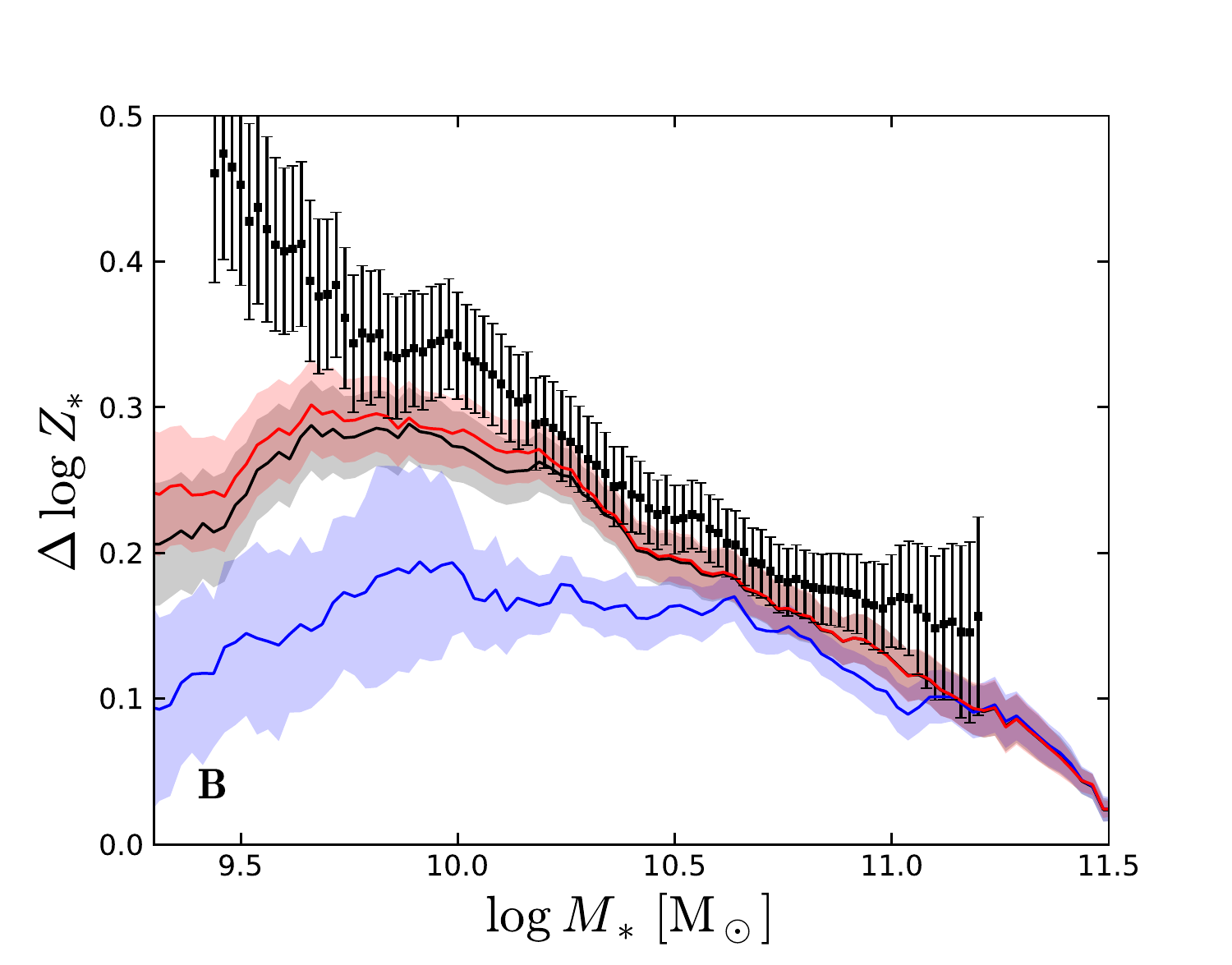} \\
\includegraphics[width=0.5\hsize]{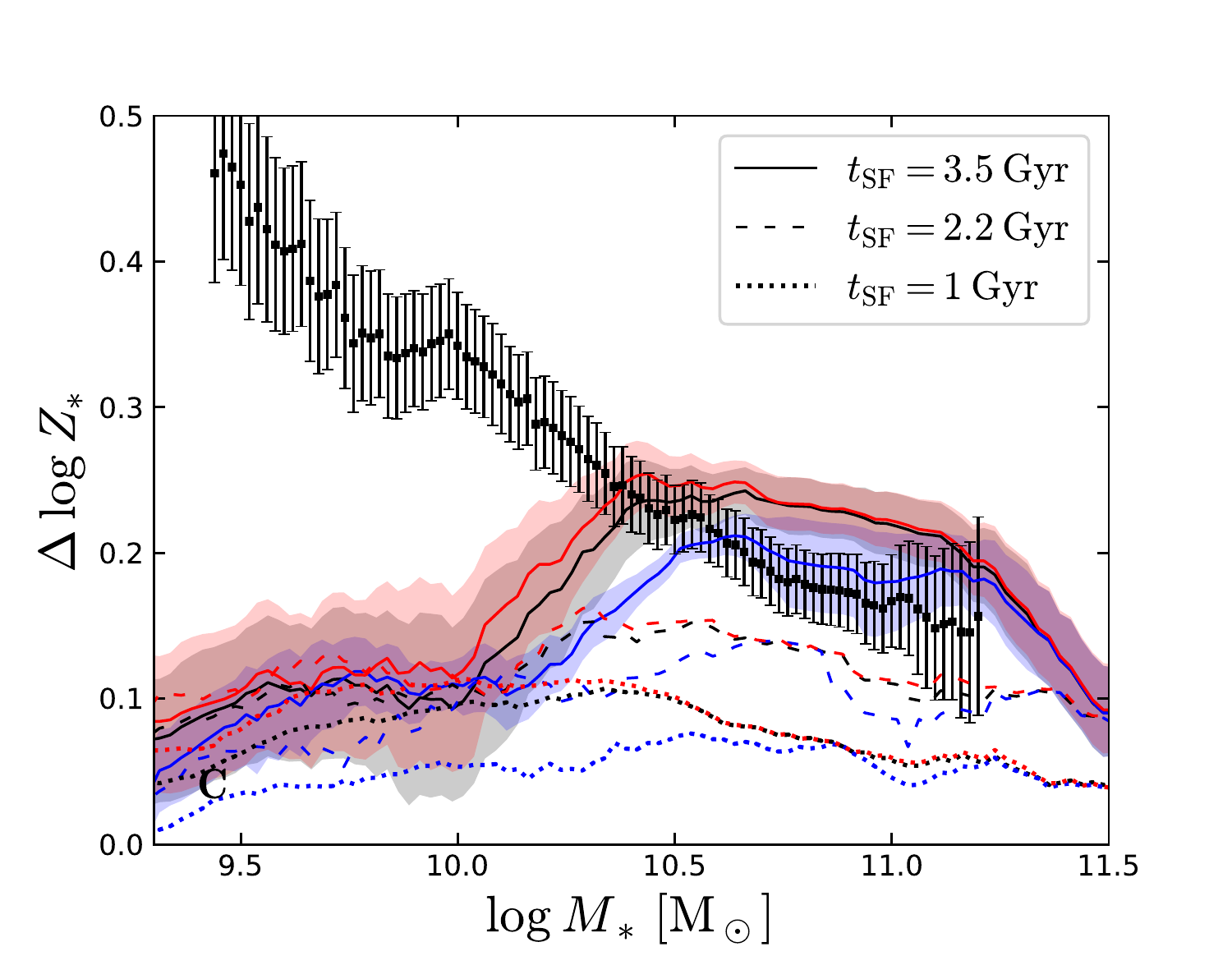}&
\includegraphics[width=0.5\hsize]{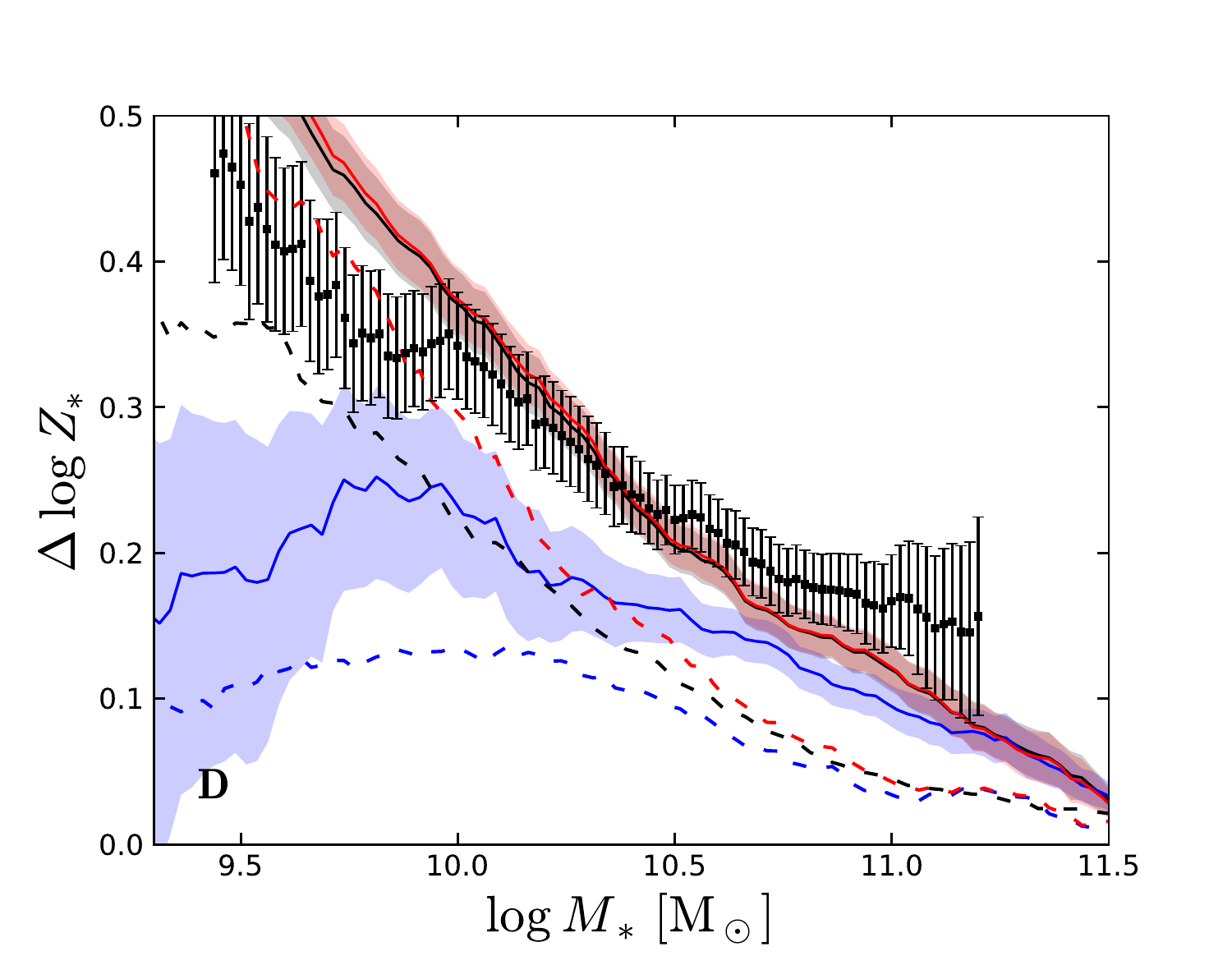} 
\end{array}$
\end{center}
\caption{Median logarithmic metallicity difference between passive and star-forming galaxies as a function of stellar mass for the four models in Table~1 (black curves: all galaxies; blue curves: central galaxies; red curves: satellite galaxies). The dashed and dotted curves in panel C are for a constant SF timescale of $t_{\rm SF}=2.2\:$Gyr and $t_{\rm SF}=1\:$Gyr, respectively. The dashed lines in panel D display the predictions of model $d$ but when pressure induced starbursts are supressed (see end of the present section). All median metallicities have been calculated using a bin of 0.025 dex in stellar mass. To smooth our functions we applied on top a sliding average of 0.2 dex. Shaded areas represent the $1 \sigma$ uncertainty on the metallicity difference. The black points show the observed metallicity differences in SDSS galaxies by T18.}
\label{ABCD}
\end{figure*}

\section{Results}

Fig.~2 shows the median logarithmic metallicity difference 
\begin{equation}
\Delta{\rm log}\,Z_*(M_*)={\rm log}{Z_*^{\rm passive}(M_*)\over Z_*^{\rm SF}(M_*)}
\label{DeltaZ}
\end{equation}
 between passive and star-forming galaxies as a function of stellar mass $M_*$,
for the four models in Table~1 (black curves) and for the observations by T18 (points with error bars), where star-forming and passive galaxies are galaxies with:
\begin{equation}
{\rm log}{{\rm SFR}\over M_\odot{\rm\,yr}^{-1}} > 0.70{\rm\,log}{M_*\over M_\odot} - 7.52
\end{equation}
and:
\begin{equation}
{\rm log}{{\rm SFR}\over M_\odot{\rm\,yr}^{-1}} <0.70{\rm \,log}{M_*\over M_\odot} - 8.02, 
\end{equation}
respectively (T18). The advantage of considering the logarithmic metallicity difference is that the dependence on $y$ through $Z_*^{\rm passive}$ and $Z_*^{\rm SF}$ cancels out, 
since both are proportional to the metal yield. 

Models $a$ and $b$ are the most standard ($a$ is the default model of C17 except for the minor modifications described in Section~2)
and both reproduce well the trend of $\Delta{\rm log}\,Z_*$ with $M_*$ at $10^{10}\,M_\odot\lsim M_*\lsim 10^{11}\,M_\odot$, although in both cases there is a systematic offset of $\sim -0.05\,$dex
of the models (black curves) with respect to the data points. 

The blue and the red curves in Fig.~2 show the $M_*$ -- $\Delta{\rm log}\,Z_*$ relation for central and satellite galaxies, separately. 
In {\sc GalICS~2.0}, passive satellite galaxies have higher stellar metallicities than passive central/field galaxies for a fixed stellar mass
(the red curves are above the blue curves). 
P15 (but not T18) find a similar trend: for a same stellar mass, passive galaxies in dense environments have higher metallicities than passive galaxies in the field.
We do not know why the results of P15 and T18 differ on this point.
However, we remark that the Universe contains more pathways to quiescence than our SAM.
Dwarf galaxies have shallow potential wells. Galactic winds can easily unbind their gas reservoir and disrupt the accretion flow onto them on scales as large as a megaparsec \citep{tollet_etal19}.
The result is stochastic SFR histories, characterised by bursts of star formation interspersed by long periods of quiescence \citep{shen_etal14}.

In {\sc GalICS~2.0}, environmental effects are the only mechanism through which a galaxy with $M_{\rm vir}\lsim 10^{12}\,M_\odot$ may become passive\footnote{In galaxies with $M_{\rm vir}> 10^{12}\,M_\odot$, quiescence is driven by the shutdown of gas accretion in massive haloes (\citealp{cattaneo_etal06}; C17).}. Passive central galaxies with $M_*\lsim 10^{11}\,M_\odot$  are usually backsplash galaxies, i.e., galaxies that have passed through a group or a cluster and come out to the other side. If we define backsplash galaxies as those that have passed within half the virial radius of the main halo \citep{Teyssier2012}, we find that $\sim8$, $\sim10$ and $\sim 19$ percent of central galaxies (with $M_*>10^{8}\,M_\odot$) within $2.5$ virial radii from host haloes with $M_{\rm h}=10^{11}-10^{12}\,M_\odot$, $10^{12}-10^{13}\,M_\odot$ and $10^{13}-10^{14}\,M_\odot$,  respectively, are backsplash galaxies, consistently with previous studies \citep{Teyssier2012,Wetzel2014}\footnote{Comparing the results of \citet{Teyssier2012} and \citet{Wetzel2014} is not straightforward because of their definition of backsplash galaxies; \citet{Wetzel2014} defines backsplash as all galaxies that remained within their host's virial radius for at
least two consecutive outputs.}.
The small number of passive central galaxies at low masses is an important element to keep in mind because it implies that the blue curves may not be statistically robust and may be the cause of the bump around $M_* \approx 10^{10} M_{\odot}$ observed in model $a$.

It is also important to realise that the metallicity of star-forming galaxies $Z_*^{\rm SF}$ and the metallicity of passive galaxies $Z_*^{\rm passive}$ that enter Eq.~(\ref{DeltaZ}) are computed in bins of $M_*$
and that the statistics behind the calculation of these two metallicities are not the same. $Z_*^{\rm SF}(M_*)$ is a very robust quantity because most galaxies with $M_*\lsim 10^{11}\,M_\odot$ are star-forming and most
of them reside in the field (hence, they are not affected by environmental effects).  $Z_*^{\rm passive}(M_*)$ is computed from a smaller population of galaxies,  most of which are satellites in the mass range of interest
 ($M_*\lsim 10^{11}\,M_\odot$).
 
 In  {\sc GalICS~2.0}, 60 -- 70$\%$ of satellite galaxies are passive. This fraction is consistent with the observational range (55 -- 95$\%$)  although the observations exhibit a passive fraction that grows with $M_*$ \citep{wetzel_etal13},
 while {\sc GalICS~2.0} finds a passive fraction independent of stellar mass, at least for  $M_*\lsim 10^{11}\,M_\odot$.
 Two effects conspire to produce this discrepancy. Strangulation may be more effective in {\sc GalICS~2.0} than in the Universe, especially in low-mass haloes, where ram pressure is not very important; hence, the passive fraction in {\sc GalICS~2.0}
 is always larger than 60$\%$. {\sc GalICS~2.0} does not contain other mechanisms to shut down star formation in galaxies with $M_*\lsim 10^{11}\,M_\odot$ besides strangulation; hence the passive fraction fails to attain values greater than 70$\%$ where it should do. 

Shutting down accretion onto satellites (model $b$) makes very little difference because haloes accrete gas only when they grow in mass and
most satellite haloes stopped growing even before the halo finder identified them as subhaloes (e.g.,T17). A freefall time after the halo stops accreting the galaxy stops accreting, too.
Shutting down accretion onto satellites simply enforces what is already happening in most cases. Hence, strangulation is not an additional physical process that our SAM needs to invoke;
strangulation arises naturally from tidal forces in the N-body simulation.

\begin{figure*}
\begin{center}
\includegraphics[width=0.95\hsize]{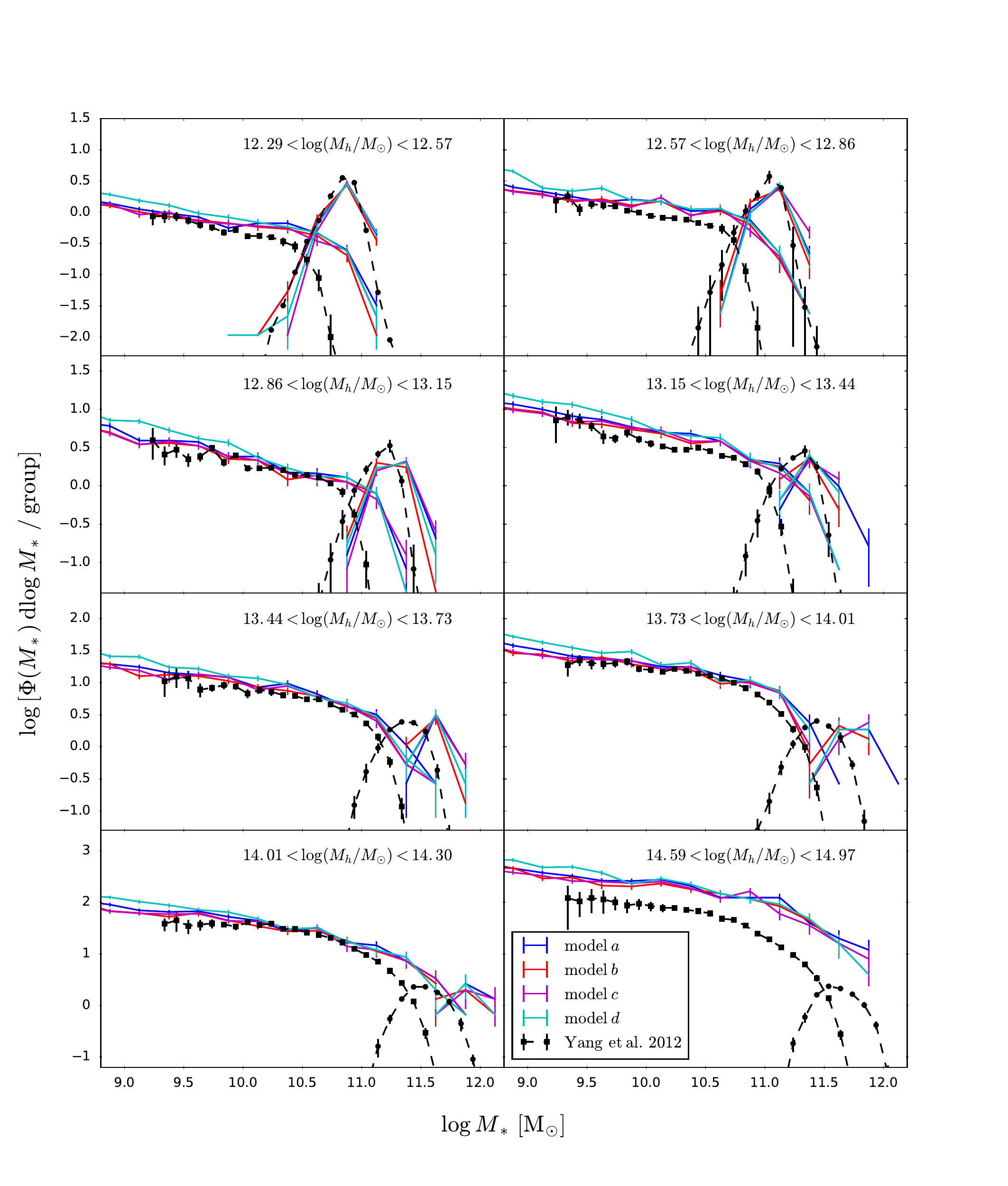}  
\end{center}
\caption{Conditional stellar mass functions for the four models in Table~1 (curves) and the observations by Yang et al. (2012; data points with error bars).
The conditional stellar mass function gives the number of galaxies with mass $M_*$ per dex of ${\rm log}\,M_*$ in a group of virial mass $M_{\rm h}$.
Squares and circles show the observations for satellite and central galaxies, respectively.}
\label{SMF}
\end{figure*}

\begin{figure*}
\begin{center}
\includegraphics[width=0.97\hsize]{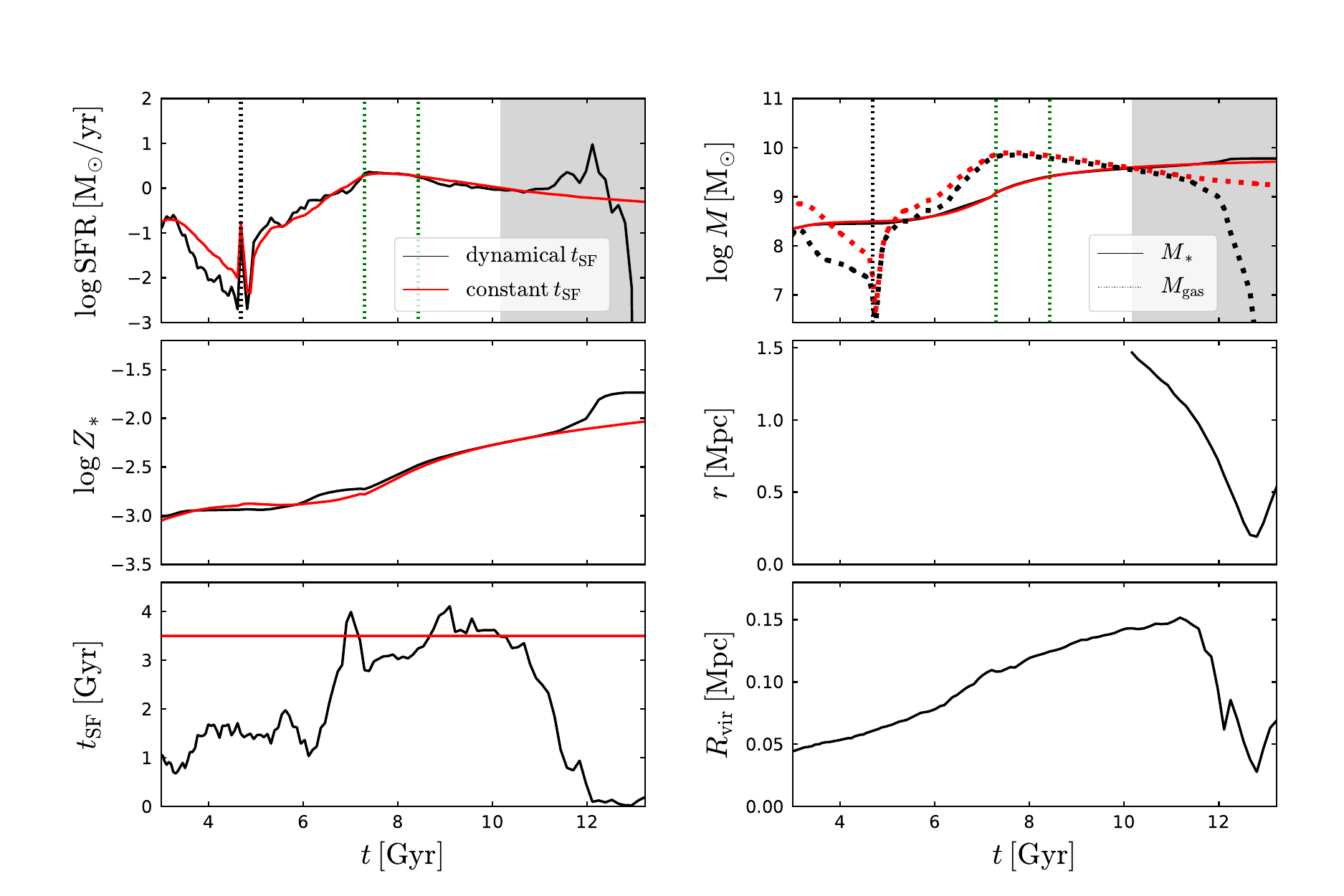}  
\end{center}
\caption{Comparison of the evolution of an individual satellite galaxy as predicted by model $b$ (black) and $c$ (red). They grey-shaded area denotes the period when the galaxy is inside the group/cluster. They vertical black and green dotted lines denote the time when the galaxy underwent a major and minor mergers respectively. In model $b$ a burst of SF, coinciding with the time of pericentric passage leads to a sharp increase of the galaxy's stellar metallicity.}
\label{713}
\end{figure*}

The change in behaviour at $M_*<10^{10}\,M_\odot$ (the black curve falls; the data points rise with an even greater slope) mirrors what we have already seen in Fig.~1: the $M_*$ -- $Z_*$ relation of \citet{Gallazzi2005} flattens below
$M_*=10^{10}\,M_\odot$, the theoretical prediction of our SAM becomes even steeper. The slope of the $M_*$ -- $Z_*$ relation is set by feedback, which becomes more important for lower-mass galaxies
because of their lower escape speeds \citep{larson74}, but the discrepancy at $M_*<10^{10}\,M_\odot$ suggests that the efficiency of feedback may increase less rapidly below a certain mass because
star formation becomes sporadic \citep{tollet_etal19}.

We note, however, that Fig.~1 considers all galaxies in our computational volume and is therefore dominated by field galaxies, which are more numerous,
while Fig.~2  is dominated by the behaviour of red galaxies, which are mainly satellites at $M_*\sim 10^{11}\,M_\odot$(see above).
Hence, our outflow rates may be correctly calibrated for field and central galaxies (the agreement in Fig.~1 is more than satisfactory) but  they may be too strong for satellite galaxies.
This possibility is not far fetched because {\sc GalICS~2.0} works on
the assumption that the mass-loading factor $\eta$ is entirely determined by the depth of the gravitational potential well  (Eq.~\ref{eta_galics}) but other physics are likely to play a role, e.g.,
the pressure of the hot gas at the centres of massive systems may suppress and confine galactic winds \citep{Schindler2005,Kapferer2009,mulchaey_tesla10}, or
it may be more difficult for satellite galaxies to attain the critical SFR density for galactic winds \citep{larson74} due to environmental effects that reduce their gas fractions (strangulation, ram-pressure stripping, tidal stripping).
A lower value of $\eta$ in satellites may explain the turning point at $M_*\approx 10^{10}\,M_\odot$ as well as the $\sim -0.05\,$dex offset of the black curves with respect to the data points in Fig.~2$a,b$.

To check to what extent Fig.~2 is affected by feedback, we have run a model in which there are no outflows in satellites (model $d$).
Fig.~2$d$ shows a dramatic increase of $\Delta{\rm log}\,Z_*$ at low masses (but not at $M_*\gsim 10^{11}\,M_\odot$) to the point that some feedback is clearly needed
(the strong effect of outflows on $\Delta{\rm log}\,Z_*$ at low masses but not at high masses had already been remarked by P15 and T18).

An independent test on the requirement for feedback in satellite galaxies comes from the conditional stellar mass function of galaxies as a function of host halo (group) mass (\citealp{Yang2012}; Fig.~3).
Fig.~3 is very similar to the one that T17 had produced with a semi-empirical model based on abundance matching rather than a SAM (the merger trees used in T17 are the same that we use here)
and this confirms that our results are robust to the choice of modelling technique. 
Both the semi-analytical and the semi-empirical approach predict an excess of massive satellites in groups with virial mass $M_{\rm h}\lsim 10^{13.4}\,M_\odot$.
Both predict too many satellites in clusters with $M_{\rm h}\gsim 10^{14.6}\,M_\odot$ (but notice that there is only one such cluster in our computational volume).
The first problem hints that many massive systems that the halo finder classifies as satellites may  have been classified as central galaxies by observers (is the Milky Way a central or a satellite galaxy?)
The second problem is most likely due to cosmic variance/poor statistics on our side.
Globally, however, the conditional stellar mass function of satellite galaxies is reproduced well by all models except model $d$,
which tends to overestimate the number of low-mass satellites in all groups.

Imposing a complete shutdown of gas accretion in satellite galaxies reduces the final stellar mass (compare the blue and the red curves in Fig.~3) but the difference is too small and the uncertainties are too large 
for Fig.~3 to be useful to discriminate between model $a$ and model $b$, which are, even in this respect, essentially indistinguishable.

Model $c$ has a constant star-formation timescale $t_{\rm SF}=3.5\,$Gyr for discs and is the one that differs most from observations (Fig.~2$c$).
We also note that model $c$ predicts a much lower fraction of passive satellites than the other three models ($<20\%$ at $10^{10}\,M_\odot\lsim M_*\lsim 10^{11}\,M_\odot$).
To find out why, we have followed the star-formation and chemical-evolution histories of a number of galaxies.
The one shown in Fig.~4 is one of them.

Fig.~4 follows the evolution of a low-mass spiral with $M_*\sim 10^{10}\,M_\odot$ in a group with $M_{\rm h}\simeq 3\times 10^{13}\,M_\odot$ (masses at $z=0$).
The black curves and the red curves show how its  SFR, gas mass, stellar mass, metallicity, distance from the centre of the group, star-formation timescale and virial radius evolve with cosmic time in models $b$ and $c$, respectively.
The only difference between the two models is the disc star-formation timescale.
We now follow its history in model $b$ and model $c$ in detail.

Just before the cosmic time $t=5\,$Gyr a major merger triggers a starburst that consumes all the gas. 
Later the galaxy regrows a disc. Hence, the gas mass and the stellar mass keep growing until $t\simeq 7.5\,$Gyr.
After $t\simeq 7.5\,$Gyr the DM halo stops growing and the galaxy ceases to accrete. Stars keep forming but the gas reservoir is now emptying.
At $t\simeq 10\,$Gyr the galaxy becomes a satellite of a  group with virial mass $M_{\rm h}\simeq 3\times 10^{13}\,M_\odot$ and its halo becomes a subhalo,
but until $t\simeq 11\,$Gyr the predictions of model $b$ and model $c$ are still quite similar. 
The difference at $t> 11\,$Gyr arises because model $b$ displays a burst of star formation that depletes the gas reservoir. The SFR peaks around $t=12\,$Gyr. 
This burst is entirely due to a sharp decrease of the disc star-formation timescale (evolution of $t_{\rm SF}$ in Fig.~4), since the last major merger and the last minor merger occurred at $t\simeq 4.7\,$Gyr and $t\simeq 8.4\,$Gyr,
respectively.
This cannot happen in model $c$, where $t_{\rm SF}=3.5\,$Gyr at all times by construction.
Model $b$ predicts a higher final stellar metallicity than model $c$ because the more stars form, the more $Z_{\rm gas}$ increases, since a larger mass of metals is deposited into a smaller gas reservoir,
and this affects the metallicities of new stellar generations.

The question is why model $b$ exhibits a sharp decrease of $t_{\rm SF}$ at $t\gsim 11\,$Gyr.
The answer is Eq.~(\ref{tSFdisc}): in model $b$, $t_{\rm SF}$ is proportional  to the disc exponential scale-length $r_{\rm d}$.
The star formation timescale $t_{\rm SF}$ decreases after $t\simeq 11\,$Gyr because $r_{\rm d}$ shrinks as the satellite approaches its orbital pericentre
(in Fig.~4, $r$ shows the distance of the galaxy from the centre of the cluster as a function of time). What causes $r_{\rm d}$ to contract, however?

In {\sc GalICS~2.0}, $r_{\rm d}$ is computed taking the disc's self-gravity into account (C17) but this is a small correction.
To the extent that the disc has a flat rotation curve with circular velocity $v_{\rm c}$, the conservation of specific angular momentum gives:
\begin{equation}
2r_{\rm d}v_{\rm c}=\lambda R_{\rm vir}v_{\rm vir},
\label{jcons}
\end{equation}
where $R_{\rm vir}$ is the virial radius of the halo (or subhalo, in the case of a satellite), $v_{\rm vir}$ is the virial velocity, and $\lambda$ is the spin parameter of the DM halo defined as in \citet{Bullock2001}.
If we further assume $v_{\rm c}\sim v_{\rm vir}$, which is inaccurate (C17) but not unreasonable, we find:
\begin{equation}
r_{\rm d}\simeq{\lambda\over 2}R_{\rm vir}.
\label{rd}
\end{equation}
Hence, $r_{\rm d}$ is essentially determined by quantities measured from the N-body simulation.
More quantitative calculations show that Eq.~(\ref{rd}) is accurate to $20\%$.

A possible complication arises because the virial radii that the halo finder (AHOP; \citealp{Tweed2009}) returns for subhaloes are in fact truncation radii. Hence, they are expected to decrease as tidal stripping peels the subhaloes' outer layers \citep{Kazantzidis2004,Diemand2007a,Diemand2007b}.
When baryonic physics are included, tidal disruption may become stronger or weaker depending on the masses of the host halo and the subhalo, and the time since the formation of the group/cluster \citep{Dolag2009, Fiacconi2016, Chua2017, Despali2017, Sawala2017, Garrison-Kimmel2017,Graus2018, Kelley2019}. For subhaloes in the mass range $10^{6}-10^9\,M_\odot$, there appears to be a consensus that their number is lower in hydrodynamic simulations with baryons relative to pure DM N-body simulations both in Milky Ways and higher-mass systems. This suggests that the contribution of baryons to the gravitational potential of the host increases the effectiveness of tidal forces \citep{Sawala2017, Chua2017, Despali2017, Garrison-Kimmel2017,  Graus2018, Kelley2019}. Things are less clear for subhaloes with masses larger than $10^9\,M_\odot$. 
\citet[{\sc eagle}]{Despali2017} found that the lack of substructures is maintained at all masses. \citet{Dolag2009} and \citeauthor{Fiacconi2016} (\citeyear{Fiacconi2016}; Ponos), found that a concentrated baryonic component makes massive substructures more resistant to tidal disruption and thus increases their number. \citet{Chua2017}, \citet{Despali2017} and \citet{Graus2018} confirmed this behaviour with the Illustris simulation but only for subhaloes with $M_{\rm vir} > 10^{10}\,M_\odot$.

Given that in the mass range considered in this article most satellites have virial masses $M_{\rm vir}>10^{10}\,M_\odot$, it is hard to quantify the effect of neglecting baryons on the disruption of substructure. However, Fig.~4 reveals a different effect: as satellites move away from the centre of the group or cluster to which they belong, their virial radii, $R_{\rm vir}$, increase again in an almost symmetric fashion to the one they decreased before the pericentric passage. This effect points to a limitation of halo finders based on three-dimensional configurations rather than six-dimensional phase-space analysis \citep{Knebe2011,Knebe2013}.
Reliance on configurations only forces AHOP to assume instantaneous tides and this approximation overestimates tidal stripping (T17).

What matters, however, is not as much $R_{\rm vir}$ as the specific angular momentum of the DM.
Disc sizes are computed by assuming that the disc has the same specific angular momentum as the DM halo (Eq.~\ref{jcons});
$R_{\rm vir}$ would not matter if the specific angular momentum of the DM were insensitive to the radius within which it is measured.
However, particles at the centre of a DM halo have lower specific angular momentum on average that particles in the outskirts.
Hence, smaller truncation radii result in smaller disc scale-lengths and shorter $t_{\rm SF}$ around the pericentre.

One may dismiss the burst of star formation in model $b$ as a numerical artifact.
However, tides may trigger bursts of star formation through other processes that our SAM neither includes nor can easily capture (i.e., tidally induced bars and compressive tides; \citealp{moss_etal98}; \citealp{renaud_etal14}).
 \citealp{lokas_etal16} used computer simulations to study the infall and evolution of disc galaxies in a cluster environment.
 Tidally induced bars formed on all orbits soon after the first pericentric passage and survived until the end of the simulations.

Tides are not the only process that can accelerate star formation around the pericentre. Ram pressure can strip gas but can also compress it and thus accelerate its conversion into stars.
Ram pressure increases with the density of the hot gas in the halo (the intracluster medium in the case of a cluster of galaxies) and with the square of the satellite's orbital speed. Both increase when approaching the pericentre.
Hence, tides and ram pressure can act together to increase the SFR around a pericentric passage.

\begin{figure}
\begin{center}
\includegraphics[width=1.\hsize]{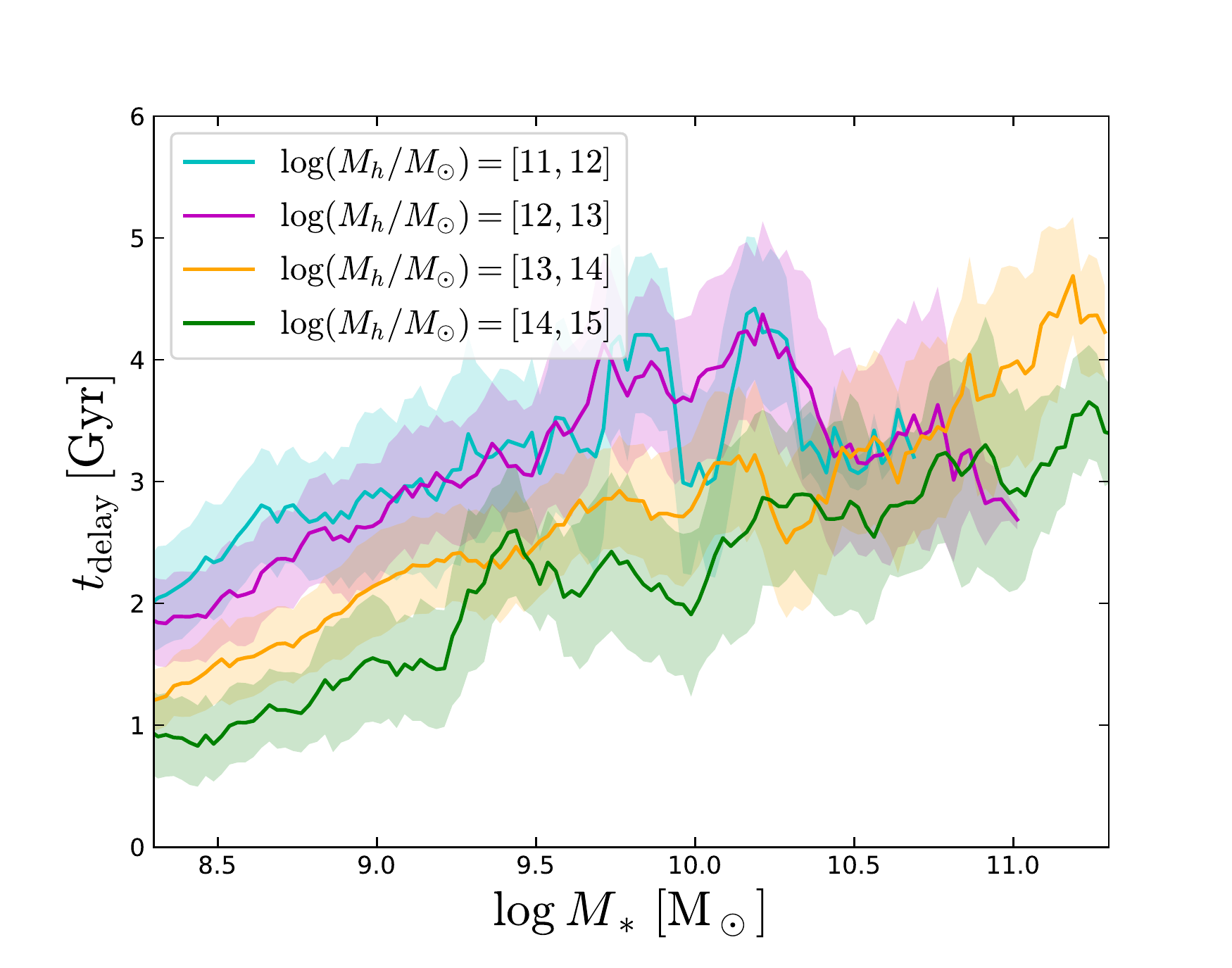} 
\end{center}
\caption{Median satellite quenching time-scale (defined as the delay between the onset of strangulation and the time a galaxy becomes passive) as a function of their current stellar mass, in bins of their current host halo virial mass at $z=0$, predicted by model $b$. Only satellites that quenched after infall and remained passive up to $z=0$ are represented. Shaded areas show the $1 \sigma$ uncertainty on the median. Overall, the quenching timescale increases with increasing stellar mass and decreasing host-halo mass. In two out of four bins a drop in the quenching timescale is observed at the high mass end.}
\label{t_delay}
\end{figure}

{\sc GalICS~2.0} does not include these physics, which are not straightforward to model, but the numerical effect discussed above produces qualitative effects similar to those that tides and ram pressure would produce if they were implemented.

Given the importance of strangulation in our model for the origin of passive satellites, it is interesting to study the delay between the onset of strangulation (the time $t_{\rm s}$ when a halo stops accreting gas) and the time when a galaxy becomes passive $t_{\rm q}$ in model $b$. We find that $t_{\rm delay}=t_{\rm q}-t_{\rm s}$ increases with $M_*$ up to $M_*\approx 10^{10.2}\,M_\odot$, then decreases with further increasing $M_*$ in two out of four $M_{\rm h}$ bins (Fig.~5). This behavior has been confirmed by numerous studies involving observations and N-body simulations (e.g., \citealp{De-Lucia2012, wetzel_etal13, Wheeler2014, Weisz2015, Wetzel2015, Fillingham2015}). Moreover, we find that satellites in more massive systems display shorter quenching timescales. Quantitatively, our predicted delay times are shorter than the values found by \citet{De-Lucia2012}, \citet{wetzel_etal13} and \citet{Wetzel2015} by $\sim 1-3\,$Gyr depending on the study\footnote{\citet{De-Lucia2012} and \citet{Wetzel2015} give values between $5-7\,$Gyr and $2-5\,$Gyr for $10^{9}<M_*/M_\odot<10^{11}$ and $10^{7.5}<M_*/M_\odot<10^{11}$, respectively. On the contrary, \citet{Wheeler2014} find a very long quenching timescale of $\sim9.5 \,$Gyr for $10^{8.5}<M_*/M_\odot<10^{9.5}$.}. This could indicate that our bursts are too strong and short-lived (due to for example their neglect of the satellites' inclination\footnote{In a physical model, the effects of ram pressure and therefore the importance of the SFR increase will depend on inclination (see Section~4).} and other parameters). Notice, however, that our delay times are based on passive satellites only, while the longer delay times found in the aforementioned studies apply to the entire satellite population and are tuned to reproduce the observed passive fraction.

\begin{figure}
\begin{center}
\includegraphics[width=1.\hsize]{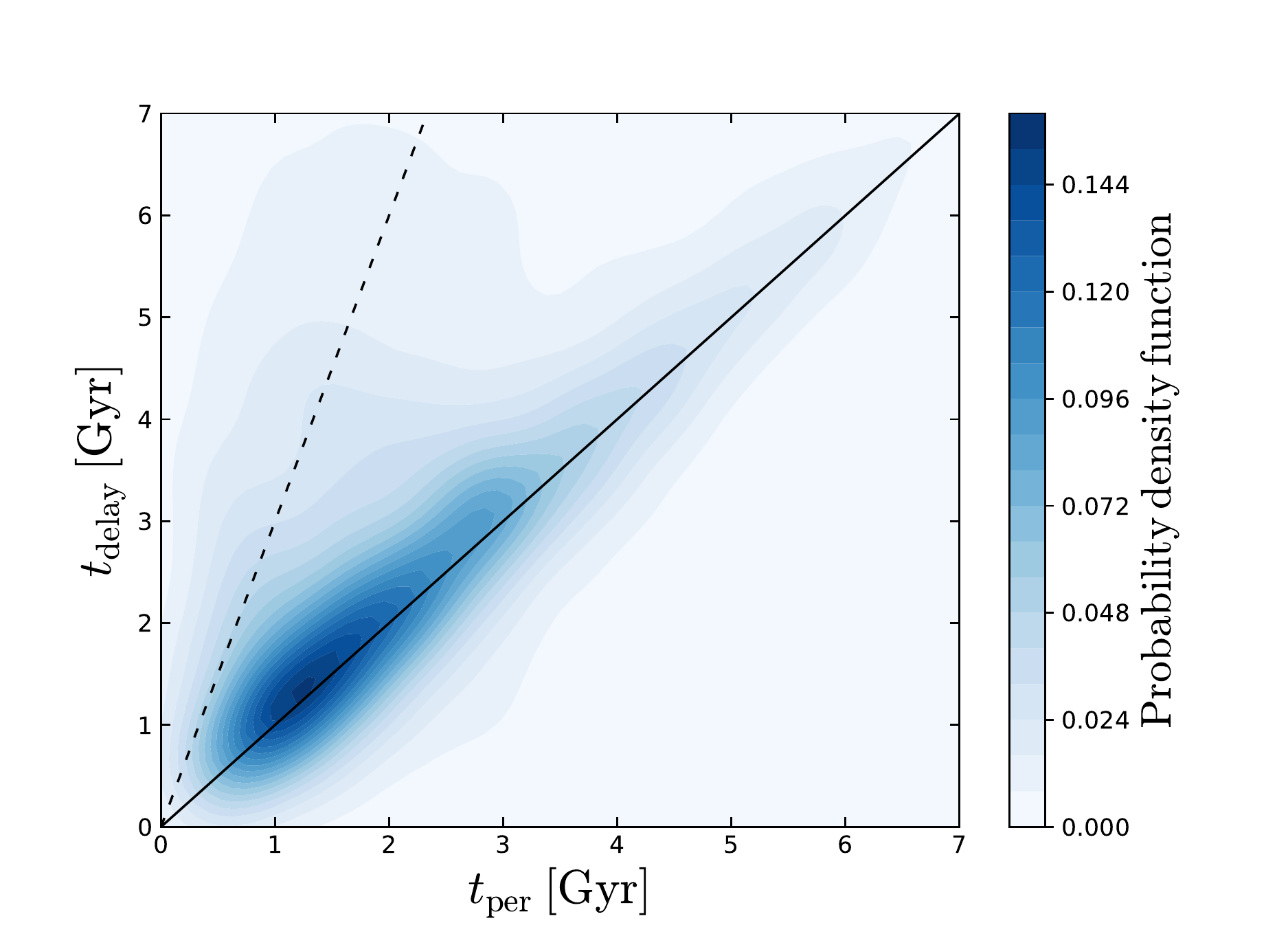} 
\end{center}
\caption{Probability density function for satellites in the $t_{\rm delay}$-$t_{\rm per}$ space, where $t_{\rm per}$ the time between the onset of strangulation and the first pericentric passage, predicted by model $b$. Most satellites cluster around the solid and the dashed lines which show $t_{\rm delay}$=$t_{\rm per}$ and $t_{\rm delay}$=$3t_{\rm per}$, respectively.}
\label{t_delay_t_per}
\end{figure}

Fig.~6 shows that most passive satellites were quenched at the first pericentric passage ($t_{\rm per}$ is the time between the onset of strangulation and the first pericentric passage). The quenching probability has a second maximum at the second pericentric passage, which occurs around 3$t_{\rm per}$ after the onset of strangulation, if we assume that it takes 2$t_{\rm per}$ for a satellite to complete an orbit around the centre of the host halo. This second maximum corresponds to satellites, in which the burst of SF at the first pericentric passage was not strong enough to exhaust all the gas.

In this article, we cannot claim to have a physical model for how bursts occur (outside mergers), but we remark that only models in which quenching is preceded by bursting produce results consistent with the observed
$M_*$ --  $\Delta{\rm log}\,Z_*$ relation. 
The model without bursts (model $c$) fails to produce enough metals in passive galaxies at low masses ($M_*\sim 10^{10}\,M_\odot$) and predicts too many metals
in passive galaxies at high masses ($M_*\sim 10^{11}\,M_\odot$). The observational trend is thus completely reversed.

Model $c$ predicts higher $\Delta{\rm log}\,Z_*$ than model $b$ at $M_*\gsim 10^{10.5}\,M_\odot$ because galaxies above $\sim 10^{10.5}\,M_\odot$ acquire a non-negligible fraction of their mass through mergers
\citep{cattaneo_etal11}, which will be very gas-rich if gas is not consumed in lower-mass objects (the final metallicity will be higher if gas is first allowed to accumulate and then exhausted in higher-mass objects, due to the anticorrelation of $\eta$ with virial mass).
Merger-driven starbursts are an effective way to form stellar populations with high metallicities, especially in massive galaxies, which have deep gravitational potential wells and behave like close boxes
(the effect is accentuated by the absence of quasar feedback in our SAM). However, we must note that only a small number of massive galaxies in model $c$ experience starbursts strong enough to move them to the red sequence resulting in a bias of their MZR towards higher metallicities.

The higher fraction of passive satellites in model $b$ ($60\%$ to $70\%$  vs. $<20\%$ in model $c$) confirms that the difference with model $c$ is due to model $b$'s more efficient star formation,
which causes many satellite galaxies to run out of gas. Using $t_{\rm SF}=2.2\,$Gyr \citep{Bigiel08} instead of $t_{\rm SF}=3.5\,$Gyr \citep{Boselli2014}\footnote{The Herschel Reference Survey used by \citet{Boselli2014} includes the Virgo cluster and may therefore be more representative of group and cluster environments, which are the primary focus of the article.
We also observed that {\sc GalICS~2.0} with the star-formation timescale from Eq.~(\ref{tSFdisc}), which gives $t_{\rm SF}=3.5\,$Gyr on average (C17), reproduces correctly gas fractions \citep{garnett02,noordermeer_etal05,zhang_etal09,Boselli2014}
and the SFR -- $M_*$ relation in the local Universe  \citep{elbaz_etal07,salim_etal07,wuyts_etal11}.} improves model $c$'s agreement
with observations but not substantially (Fig.~2C; dashed curves). The fraction of passive galaxies is higher but still remains below $\sim 30-40\%$ over the whole mass range and the metallicities of low-mass galaxies
($M_*\sim 10^{10}\,M_\odot$) are still underpredicted. The only significant difference of using $t_{\rm SF}=2.2\,$Gyr is that the metallicities of massive galaxies ($M_*\gsim 10^{11}\,M_\odot$)
are no longer overpredicted.

Lowering the value of $t_{\rm SF}$ in model $c$ even further increases the fraction of passive galaxies but does not improve the agreement with the observed metallicities (Fig.~2C; dotted curves). For shorter $t_{\rm SF}$ galaxies in a leaky-box model exhaust their gas reservoir and reach their maximum stellar metallicity:
\begin{equation}
\label{Zstar}
Z_{*,\rm max} =  {1-R\over 1-R+\eta}y
\end{equation} 
faster (Eqs.~\ref{Mstars_evol},~\ref{MstarsZ_evol}). However, this leads to gas-poorer mergers that prevent the metallicities of massive galaxies, which assembled most of their mass through mergers, from growing. Moreover, in a model with lower $t_{\rm SF}$ star-forming galaxies reach a given mass faster, without having the time to dilute their gas reservoir significantly by the pristine infalling gas (their MZR is elevated). Counterintuitively, we, therefore, find the difference $\Delta{\rm log}\,Z_*$ between passive and star-forming galaxies as a function of mass to be smaller when the constant $t_{\rm SF}$ is decreased. In contrast, the success of model $b$ relies on decreasing only the SF timescale of satellite galaxes (SF bursts), while at the same time ensuring that high-mass galaxies have acquired enough gas prior to the onset of strangulation ($t_{\rm SF}$ decreases only after the entry time).

In order to disentangle the effects of pressure-induced starbursts and the suppresion of outflows, we consider a variant of the closed-box model $d$ (Fig.~2d, dotted lines), in which we force the disc scale-length $r_{\rm d}$ of satellite galaxies to remain constant after the entry in the group/cluster, so that there are no changes in the SF timescale. While at the lower mass end, satellites manage to reach and even surpass the observed $\Delta {\rm log} Z_*$, bursts are necessary for satellites with masses higher than $M_* \simeq 10^{10} M_{\odot}$ to exhaust their gas and enrich up to the observed metallicities.

\section{Effects of stripping}
In the previous section, we explored models that included strangulation (the shutdown of gas accretion onto satellites) but did not include any stripping.
We now wish to explore how ram pressure and tidal stripping could affect the metallicities of satellite galaxies. 

Ram pressure strips gas more effectively at large radii \citep{Bekki2014,Steinhauser2016}
in agreement with observations that cluster galaxies have more concentrated star formation than their field counterparts \citep{Bretherton2013}.
 Truncating discs is equivalent to removing the gas with the lowest metallicities and increases the mean metallicity of the gas that is left over to form stars.
 \citet[Illustris]{Genel2016} and \citet[{\sc eagle}]{Bahe2017} found that this mechanism accounts for a significant part of the metallicity enhancement of the gas in satellites compared to field galaxies. In a similar manner, tidal stripping of outer, metal-poor stars can simultaneously increase the average stellar metallicity and reduce the stellar mass of satellite galaxies, although this is likely to be a minor effect \citep{Pasquali2010}.

To assess the maximum effect that these mechanisms may have on stellar metallicities we have explored an extreme model in which stripping removes gas but does not remove any metals from galaxies. For this calculation we have computed $t_{\rm SF}$ with Eq.~(2) but we have assumed that the disc scale-length of satellite galaxies remains constant after the entry time. Otherwise,  the artificial contraction of gaseous discs brings the gas so close to the centre that its stripping is artificially impossible.

Following \citet{Gunn1972}, the ram pressure exerted by the ambient hot gas on the disc of a satellite galaxy is equal to:
 \begin{equation}
 P_{\rm ram}(r) = \rho_{\rm h}(r)\upsilon_{\rm sat}^2(r){\rm cos}\:\theta ,
 \end{equation}
where $\rho_{\rm h}(r)$ is the density of the intracluster medium, $\upsilon_{\rm sat}(r)$ is the velocity of the satellite in the host's reference frame ($r$ is the distance of the satellite from the centre of the host) and $\theta$ is the inclination angle between the disc's rotation axis and its velocity (the wind direction). Here, $\upsilon_{\rm sat}(r)$ is determined from the N-body simulation and we assume $\theta = 0$ (all satellites move face-on) to maximize ram-pressure stripping. The hot gas density is computed assuming a profile of the form:
\begin{equation}
\label{Faltenbacher}
\rho_{\rm h}(r) = \frac{\rho_{\rm h,0}}{\left(1+\frac{cr}{\alpha R_{\rm vir}}\right)^3}
\end{equation}
such as the one that is found in adiabatic simulations (Andreas Faltenbacher, private communication). Here, $\rho_{\rm h,0}$ is the central hot gas density, $R_{\rm vir}$ is the virial radius, $c$ is the concentration parameter of the DM halo and $\alpha$ is a parameter that sets the ratio between the core radii of the hot gas and the DM. The values of $\rho_{\rm h,0}$ and $\alpha$ are determined by the conditions: 
\begin{equation}
\begin{split}
M_{\rm hot} & = 4\pi \int_0^{R_{\rm vir}} \rho_{\rm h}(r) r^2 {\rm d}r = 4 \pi \rho_{\rm h,0} \: r_{\rm c,h}^3 \int_0^{\frac{c}{\alpha}} \frac{x^2}{(1+x)^3} {\rm d}x \\
&=4 \pi \rho_{\rm h,0} \: r_{\rm c,h}^3 \bigg(\frac{4c/\alpha+3}{2(c/\alpha+1)^2}+{\rm log}\Big(\frac{c}{\alpha}+1 \Big)-\frac{3}{2} \bigg)
\end{split}
\end{equation}
and 
\begin{equation}
\lim_{r\to\infty} \frac{\rho_{\rm h}}{\rho_{\rm NFW}} = \frac{ M_{\rm hot}}{M_{\rm vir}}.
\end{equation}

\begin{figure}
\begin{center}
\includegraphics[width=1.\hsize]{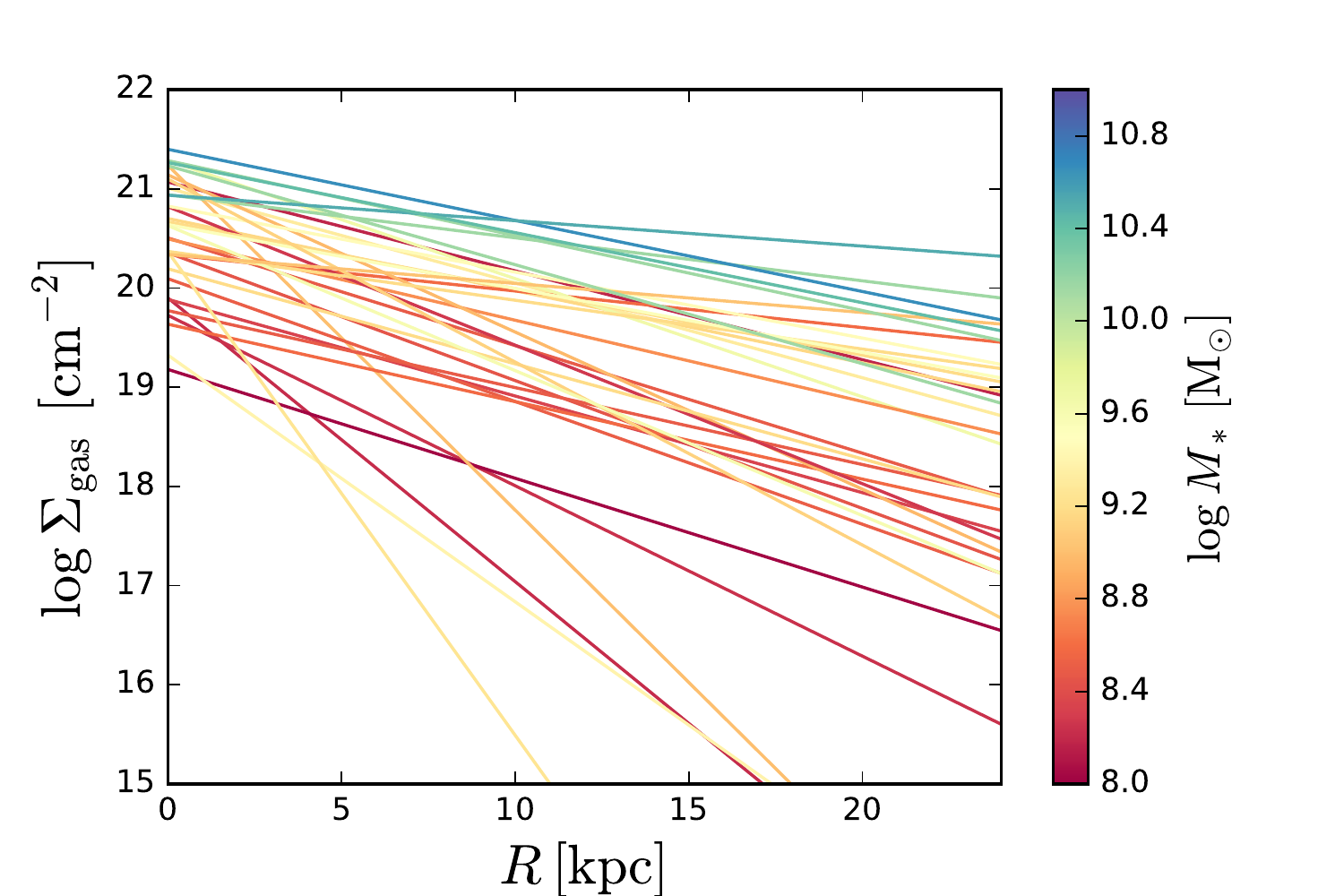} 
\end{center}
\caption{Gas surface-density profiles for discs in {\sc GalICS~2.0}, color-coded according to the stellar mass of the galaxy. The profiles are computed assuming that the exponential scale-length of the gaseous disc is twice that of the stellar disc ($f_{\rm d}=2$).}
\label{NH_R}
\end{figure}

The gravitational restoring force per unit area on the gas exerted by the stars and the gas in the disc at radius $R$ is given by:
\begin{equation} \label{restore}
\begin{split}
P_{\rm rest}(R) & = 2 \pi G \Sigma_{\rm disc}(R) \Sigma_{\rm gas}(R) \\
& = 2 \pi G  \Big[\Sigma_*(0) e^{-\frac{R}{r_{\rm d}}}+\Sigma_{\rm gas}(0) e^{-\frac{R}{f_{\rm d}r_{\rm d}}}\Big]\Sigma_{\rm gas}(0) e^{-\frac{R}{f_{\rm d}r_{\rm d}}},
\end{split}
\end{equation}
where $\Sigma_{\rm disc}(R)$ is the surface density of the disc at radius $R$, $\Sigma_*(0)$ and $\Sigma_{\rm gas}(0)$ are the central surface densities of the stellar and the gaseous disc, respectively, $r_{\rm d}$ is the exponential scale-length of the stellar component and $f_{\rm d}$ is the factor by which the gaseous disc is larger than the stellar one. The profiles found in {\sc GalICS~2.0} for $f_{\rm d}=2$ (Fig.~7) compare well with those in \citet{Fillingham2016} who compiled observational gas surface density profiles from the THINGS, Little THINGS and SHIELD data sets \citep{Walter2008,Cannon2011,Hunter2012,
McNichols2016,Teich2016}. We have also considered $f_{\rm d}=3$ to assess the stronger impact that stripping could have on more extended discs.

Eq.~(\ref{restore}) neglects the contribution of the DM halo to the gravitational potential (it is possible for a gaseous mass to escape the disc but be restrained by the halo's gravitational force) and therefore contributes to maximizing the effects of ram-pressure.

At each timestep of the satellite's orbit we evaluate the radius $R_{\rm strip}$ for which $P_{\rm ram}(r) = P_{\rm rest}(R_{\rm strip})$ and the gas mass that is stripped from the satellite:
\begin{equation}
M_{\rm gas,stripped}=2 \pi \int_{R_{\rm strip}}^{R_{\rm t}} \Sigma_{\rm gas}(r)r \rm{d}r 
\end{equation} 
after which we set the truncation radius $R_{\rm t} = R_{\rm strip}$ if $R_{\rm strip}<R_{\rm t}$. The stripped material is added to the hot gas halo of the host, which is not allowed to cool in {\sc GalICS~2.0}. 

Tidal stripping of stars and gas is computed at each pericentric passage as in T17. Differently from ram-pressure stripping, tidal stripping does not result in a truncation radius in our model, since we assume that between two pericentric passages the disc has enough time to relax back to an exponential profile (in fact, many stripped discs exhibit an antitruncated profile; \citealp{Pranger2017}).

Central galaxies are affected by stripping indirectly because they grow by merging with satellites, the properties of which have been modified by stripping.

\begin{figure}
\begin{center}
\includegraphics[width=1.\hsize]{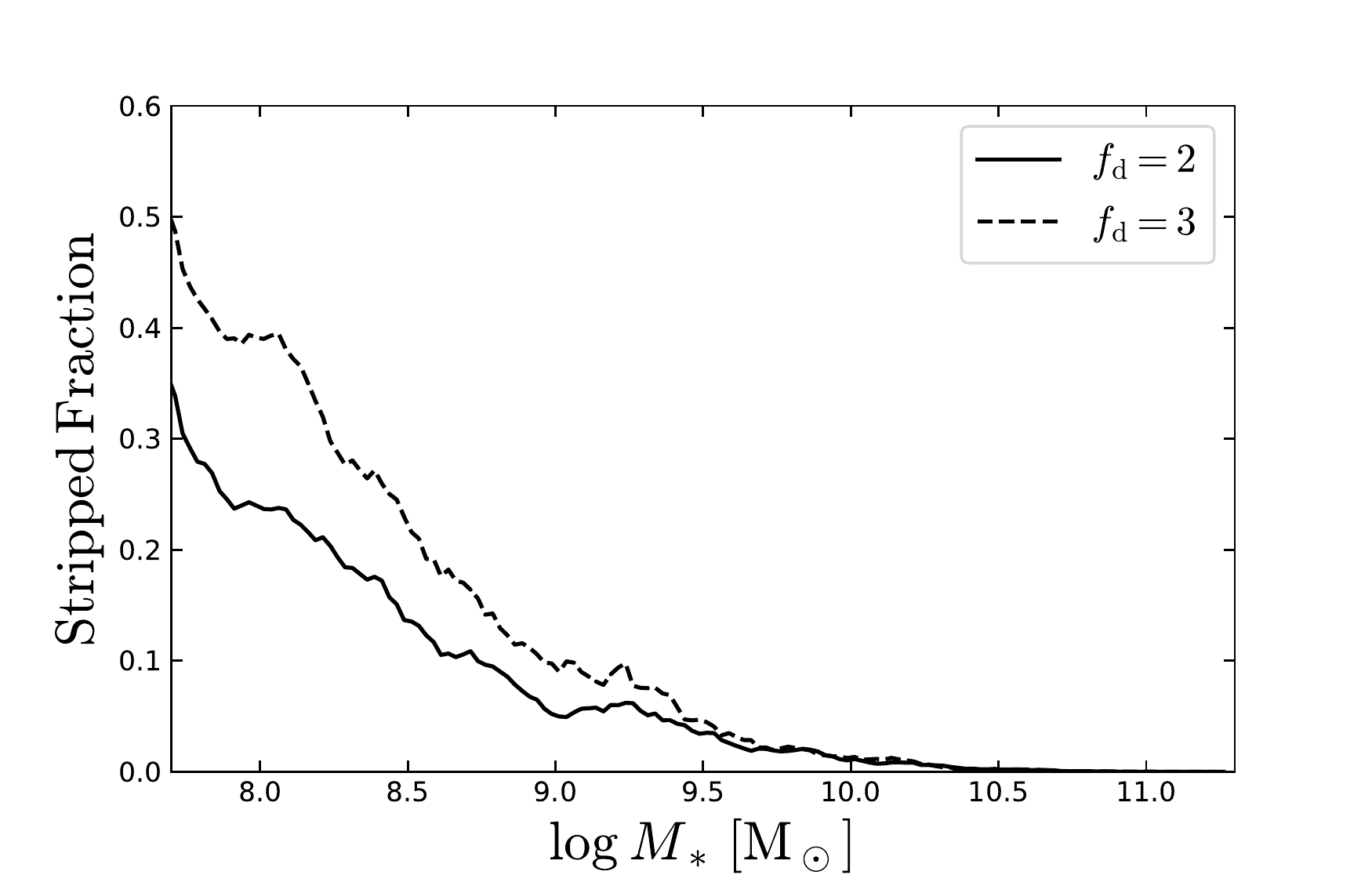} 
\end{center}
\caption{Median stripped gas fraction as a function of stellar mass. Solid and dashed curves show models where $f_{\rm d}=2$ and $f_{\rm d}=3$, respectively.}
\label{gas_stripped}
\end{figure}

Fig.~8 shows the median gas fraction stripped from satellites due to the combined effect of ram-pressure and tides for $f_{\rm d}=2$ (solid curve) and $f_{\rm d}=3$ (dashed curve). The stripped fraction is the stripped gas mass over a galaxy's entire history divided by the sum of the stripped mass and the mass of the gas in the galaxy at $z \sim 0$. 

In the mass range of interest ($M_* \gsim 10^{9.5}\: M_{\odot}$), the median stripped fraction is less than $0.05$ for both $f_{\rm d}=2$ and $f_{\rm d}=3$. Therefore, one expects a maximum metallicity increase of ${\rm log}(1/0.95) \simeq 0.02 \: {\rm dex}$. 

In reality, the metallicity of the gas can increase by more than that if a galaxy acquires most of its gas through a merger with a lower mass galaxy, which has a higher metallicity enhancement due to its higher stripped fraction. However, stellar metallicities are much less sensitive to this effect because variations in $Z_*$ are never as large as the variations in $Z_{\rm gas}$ that drive them. Moreover, the stripped fraction for stars is always lower than the one estimated for gas in Fig.~8 because stars are affected only by tidal stripping and have a smaller scale radius.

\section{Discussion}

We have used the {\sc GalICS}~2.0 SAM to investigate metal enrichment and the mechanisms that quench star formation in satellite galaxies. 
In {\sc GalICS~2.0}, satellite galaxies cease to accrete (strangulation) as soon as they enter the virial radius of their host system and often even earlier, but
quenching does not occur until a burst of star formation uses up most of the gas around the first pericentric passage.
Only models where quenching is preceded by bursting are consistent with the high fraction of red satellites observed by \citet{wetzel_etal13} and
the metallicity enhancements measured by P15 and T18 in passive galaxies.

The strength of our work is that orbits are derived from cosmological N-body simulation and mass-loading factors are calibrated to reproduce the stellar mass function of galaxies.
Neither is tuned to reproduce the difference in metallicity between passive and star-forming galaxies. The weakness is that {\sc GalICS~2.0} lacks a physical model for how
tidal bars, tidal compression and ram pressure affect the SFR of galaxies.
We have demonstrated that bursting and quenching are necessary to reproduce the observed properties of satellite galaxies.
We have not proven their necessity on physical grounds.
We can strengthen our case, however, by referring to hydrodynamic simulations and observations that show evidence for bursting and quenching.

Hydrodynamic simulations  have demonstrated that ram-pressure stripping (RPS) can remove gas from satellite galaxies
\citep{Mayer2006, Kronberger2008,Kapferer2009,Nichols2015,Emerick2016,Steinhauser2016,Simpson2018,Hausammann2019}.
The predominant role of RPS over tidal stripping is widely recognised (e.g., \citealp{Simpson2018}).
\citet{Mayer2006} noted that  gravitational tides can aid RPS by diminishing the overall gravitational potential of satellites but tides also induce bar formation, which funnels gas inwards.
Their point was that this would make the gas harder to strip but we remark that, by  funnelling gas inwards, tides also increase its SFR.

The effectiveness of RPS as a quenching mechanism is a strong function of satellite mass and pericentric radius \citep{Mayer2006,Fillingham2016,Simpson2018}.
The Auriga cosmological hydrodynamic simulation finds that, in a system similar to the Local Group, 
RPS can quench star formation in dwarf galaxies but only below a stellar mass  $M_*\sim 10^7\,M_\odot$ \citep{Simpson2018}.
A semi-analytic calculation based on the {\sc Hi} surface-density profiles of 66 nearby galaxies finds that RPS becomes important at satellite stellar masses below $\sim 10^8\,M_\odot$,
and even then is not fully effective in most systems unless the density of the hot circumgalactic medium is higher than the canonical value $10^{-3.5}{\rm\,cm}^{-3}$ \citep{Fillingham2016}.
High-resolution wind-tunnel simulations \citep{Emerick2016,Hausammann2019} confirm that RPS can remove  the hot diffuse gas very efficiently but find that cold gas ($T<10^3\,$K) is remarkably resilient\footnote{\citet{Hausammann2019} argue that \citet{Mayer2006} and \citet{Simpson2018}  found evidence for quenching 
simply because they did not let the gas cool below $10^4\,$K and condense into molecular clouds.}.

All the above studies are for systems on the mass-scale of the Local Group but Steinhauser et al. (2016; {\sc arepo}) found similar results for satellites in a cluster environment.
They confirmed that cold gas is very difficult to strip unless the pericentric radius is very small but they also noted that inefficient stripping is balanced by increased star formation in discs compressed by ram pressure.

The conclusion of this first set of studies is that RPS is an effective mechanism for strangulation but not for quenching, at least in the mass range that is relevant for this article ($M_*>10^{9.5}\,M_\odot$).
This is good news for our SAM, which neither includes RPS nor requires any strangulation mechanism.
In {\sc GalICS~2.0}, galaxies cease to grow within a freefall time from when their haloes cease to do so. Hence, there is virtually no accretion onto satellite galaxies.

Having discussed ram pressure in relation to strangulation and quenching, we now wish to examine its role as a possible trigger for starbursts.
\citet{Bekki2003} ran wind-tunnel simulations of a molecular cloud moving through the hot gas at the centre of the Coma cluster 
and found that $80\%$ of the gas was converted into stars in just over $10\,$Myr.
Simulations of satellite galaxies moving through a cluster environment confirmed their result by showing that infall into a cluster increases the SFR by a factor of three
\citep{Kronberger2008} or even by up to an order magnitude depending on the density of the intracluster medium \citep{Kapferer2009}.

More recently, \citet{Bekki2014} run a parameter study in which he examined the evolution of the SFR as a function of satellite mass (he considered three galaxies with masses from a dwarf irregular to the Milky Way),
host-halo mass (he explored the mass range from $M_{\rm h}=10^{13}\,M_\odot$ to $M_{\rm h}= 10^{15}\,M_\odot$), orbital parameters (hence, the pericentric radius $r_{\rm p}$), and inclination.
All galaxies exhibit decreasing SFRs until they reach the pericentre.
Then, star formation may be enhanced, reduced, quenched, or unaffected, in which case it continues its gentle decline.
The least affected galaxies are edge-on discs (the effects of ram pressure are strongest when the velocity and the angular momentum are aligned and weakest when their orientations are orthogonal).
$M_*/M_{\rm h}$ is the key quantity  that determines the probability of quenched vs. enhanced star formation.
Quenched or reduced star formation is typical of dwarves in cluster environments.
For massive spirals, enhanced star formation is far more common. 
The enhancement factor can approach an order of magnitude in the case of a face-on disc \citep{Bekki2014} and occurs because the molecular fraction increases with pressure \citep{Henderson2016}.

Measuring a temporary increase of the SFR at the first pericentric passage does not solve the problem of the long-term evolution of star formation in satellite galaxies.
\citet{Bekki2014} showed that star formation may be enhanced at the first pericentric passage and reduced at the second, or that it may be unaffected at the first pericentric passage and enhanced at the second.
Bursting followed by quenching is one evolutionary path. There is no reason why it should be universal, but our results do not require that.
All we need to explain the metallicity difference between passive and star-forming galaxies is that passive galaxies experienced a phase of enhanced star formation between the onset 
of strangulation (the time they ceased to accrete gas) and the time of quenching (the time they ceased to form stars).
The work by \citet{Kronberger2008}, \citet{Kapferer2009} and \citet{Bekki2014} shows that this scenario is consistent with what we learn from hydrodynamic simulations.
 
 From an observational standpoint, there is little evidence for star formation at the centres of local clusters but the presence of a conspicuous population of star-forming galaxies in the cores of rich clusters at $z\sim0.5$
 has been known since \citet{Butcher1984}.  \citet{Finn2005}, alike, found a significant population of starburst cluster galaxies at $z\simeq0.75$  using the ESO Distant Cluster Survey (EDisCS) 
 and remarked 
 that these galaxies can be matched to poststarburst galaxies  (PSG) at $z\simeq0.45$, assuming the poststarburst phase lasted five times longer than the starburst phase. \citet{Paccagnella2017} investigated the occurence and properties of PSGs\footnote{PSGs are thought to result from fast quenching mechanisms and exhibit properties intermediate between passive and star-forming galaxies for $\sim1\,$Gyr after the halt of SF.} galaxies in 32 clusters in the local Universe ($0.04<z<0.07$)
 and found that the fraction of PSGs is similar to the fraction of galaxies with slowly decreasing SFRs (SRFs that decrease on timescales longer than $2\,$Gyr).
 As in \citet{Paccagnella2017} the post-starburst phase lasts for $1\,$Gyr by construction, 
 this finding implies that two thirds of passive galaxies reached the red sequence through the post-starburst route.
 
On the opposing side are \citet{Treu2003}, who observed the morphological distribution in a cluster with $R_{\rm vir}=1.7\,$Mpc at redshift $z=0.4$ out to radius $r=5\,$Mpc.
 They argued that tidal/ram-pressure triggering of star formation and RPS should not be important at $r>0.8\,$Mpc;
 hence,  they concluded, these processes cannot explain the gradual increase of the early-type fraction from $40\%$ in the field ($r>5\,$Mpc) to $70\%$ in the central region.
 Slow strangulation and harassment were proposed as more likely mechanisms.
 We note, however, that \citet{Treu2003}'s early-type fractions at $r\sim 0.6\,$Mpc and $r\sim 3.7\,$Mpc ($50\%$ and $45\%$, respectively) are indistinguishable within the error bars
 and that  $r\sim 3.7\,$Mpc is the outermost radius for which  \citet{Treu2003} have data for the cluster in question. Moreover, a double mechanism and/or timescale scenario is plausible, in which SF is quenched rapidly by a process that acts primarily on the gas content (RPS or RP induced bursts), while morphological transformation occurs on longer timescales \citep{Poggianti1999,Paccagnella2017}.

Considering the dependence of the passive fraction not only on radius $r$ but also on radial velocity $v_r$ lead  \citet{mahajan_etal11} to an opposite conclusion to \citet{Treu2003}.
The radial velocity was used to separate infalling galaxies, which have not reached their pericentres yet, virialised galaxies, which may have done several orbits, and backsplash galaxies,
which are likely to have experienced only one pericentric passage.
 \citet{mahajan_etal11}  found  that the fraction of galaxies with ongoing or recent star formation is higher in the infalling population than in the other two, which contain similar fractions of galaxies with ongoing or recent star formation:
 star-forming galaxies that fall into a cluster keep forming stars all the way down to the centre but one pericentric passage is enough to quench their star-formation activity.
 
 Slow strangulation predicts a broad unimodal of galaxy colour. Rapid quenching predicts a blue peak, a red peak, and very few galaxies in the green valley that separates them.
 The colour distribution of satellites in clusters is bimodal but the green valley is not empty. The best fit is for a delayed-quenching scenario (slow strangulation followed by rapid quenching; \citealp{wetzel_etal13}).
 The phase-space distribution and the stellar spectra of star-forming, recently quenched and quiescent galaxies support the notion of  strangulation over a $2.3\,$Gyr period followed by quenching $0.4\,$Gyr after the first pericentric passage
 (\citealp{muzzin_etal14}; recently quenched galaxies tend to exhibit low $r$ and high $|v_r|$; furthermore positive $v_r$ are more common than negative ones, suggesting that quenching is more likely to occur after a pericentric passage).
 
 The difference between our scenario and the one proposed by  \citet{wetzel_etal13} and  \citet{muzzin_etal14} is that quenching is due to bursting rather than gas removal.
 Our interpretation is in better agreement with the metallicities of quenched satellites and with hydrodynamic simulations, which have shown how difficult it is for ram pressure to strip all the gas.
 The only problem is the relatively strong calcium K absorption line that \citet{muzzin_etal14} observe in the spectra of recently quenched galaxies,
 which cannot be reproduced by a very young stellar population. 
 We expect, however, the final burst to occur after strangulation has depleted the gas reservoir considerably or for galaxies to continue forming stars and quench a significant time after the burst in case of less prominent starbursts (Fig.~4 shows one of the strongest starbursts to illustrate the phenomenon more clearly
 but is exceptional in regard of the SFR that is attained).
 Hence, spectral modelling would be needed to investigate the problem's quantitative significance.
 
 The metallicities of star-forming satellites are another diagnostic of how the metallicity difference between passive and star-forming galaxies originates.
 If the difference builds up gradually, the metallicities of star-forming satellites should be intermediate between those of passive galaxies and  star-forming galaxies in the field, and should increase at low $r$ for a given stellar mass.
 If the final burst makes an important contribution to $\Delta{\rm log}\,Z_*$, then the metallicities of star-forming satellites should be closer to those of star-forming field galaxies with the same stellar mass than to the metallicities of passive galaxies.
 \citet{Petropoulou2011} found that, for $M_*\gsim 10^{9.5}\,M_\odot$, the $M_*$ -- $Z_{\rm gas}$ relation in the Hercules cluster is consistent with \citet{amorin_etal10}'s MPA/JHU reference sample of star-forming galaxies.
 Furthermore, observations in the nearby clusters A1656 (Coma),  A1367, A779 and A634 \citep{petropoulou_etal12} have shown that the mean gas-phase metallicities
 of star-forming galaxies at $r\le r_{200}$ and $r>r_{200}$ are the same at all masses
 in  A779 and A634, and differ very little in A1367. Only in Coma, the most massive of the four clusters, and only for star-forming galaxies with $M_*< 10^{9.5}\,M_\odot$
 is the metallicity at $r\le r_{200}$ clearly larger than the metallicity at $r>r_{200}$ (only dwarves in clusters with $M_{\rm h}\gsim 10^{15}\,M_\odot$ experience significant increases of $Z_*$ while they are still star-forming and have not approached
 their pericentres yet). 
 
Using the SDSS DR4 catalogue,  \citet{Pasquali2012} confirmed this result and showed that the gas-phase metallicities of star-forming satellites with $M_*>10^{9.5}\,\rm{M_{\odot}}$ are enhanced by less than $\sim0.025\,$dex compared to their field counterparts. We expect the difference in stellar metallicity between star-forming central and satellite galaxies to be even less pronounced, since $Z_{\rm gas}$ reflects the current state of a galaxy while $Z_*$ represents an average over its entire SF history.

Interestingly, a pure chemical evolution study by \citet{Spitoni2017}, which considered neither the cosmological context nor the environment, came up with a similar conclusion that shorter SF timescales are needed to explain the metallicity of passive galaxies. 
 
Even with bursting, there is still a small offset between the $\Delta{\rm log}\,Z_*$ predicted by {\sc GalICS~2.0} and those measured by T18.
 The mechanism that causes the bursts is irrelevant in this regard as long as the final gas mass is virtually zero. Stripping the metal poor gas residing in the outskirts of satellites, does result into higher gas-phase metallicities as \citet{Genel2016} and \citet{Bahe2017} have proposed, but the effect on stellar metallicities is not large enough to resolve this discrepancy.
Moreover, assuming that satellite galaxies have lower mass-loading factors than field galaxies with the same mass (\citealp{Schindler2005,Kapferer2009,mulchaey_tesla10}; and the discussion in Section~3 of this article)
can solve the problem at  $M_*\lsim 10^{10.3}\,M_\odot$ without affecting the stellar mass function significantly, but cannot remove the offset at larger masses.

Without a physical model for starbursts, one could imagine that a slight underprediction of their strength is the reason for the offset at the high mass end. However, our purpose here was not to constrain the strength and dependencies of the pressure induced bursts but to show their necessity for reproducing both the mass distribution and the MZR relation
of passive satellite galaxies. 

Another caveat is that the SF timescale in {\sc GalICS~2.0} does not depend on redshift. Observations out to redshift $z=2.5$ have shown that the SF timescale for molecular gas decreases with redshift as $t_{\rm SF,mol} \propto (1+z)^{-\alpha}$, with $0.3 \lesssim \alpha \lesssim 1$ \citep{Tacconi2013,Santini2014,Genzel2015,
Scoville2017,Tacconi2018}\footnote{At higher redshifts this trend appears to flatten \citep{Bethermi2015,Schinnerer2016}.} and that the molecular gas fraction at a given stellar mass increases with redshift (\citealp{Daddi2010,Tacconi2013,Tacconi2018}; see also the theoretical predictions of \citealp{Lagos2011}, \citealp{Lagos2015} and \citealp{Popping2014}). By taking these observations into account and assuming that in passive galaxies strangulation started at a look-back time equal to their mean mass-weighted ages (ranging from $\sim 7$ Gyr ago for $M_*=10^{9.4} M_{\odot}$ to $\sim 11$ Gyr ago for $M_*=10^{11.2} M_{\odot}$), T18 managed to make their passive galaxies acquire the observed metallicities relatively fast (in just $\sim$2 Gyr for a closed-box and $\sim$5.5 Gyr for a leaky-box model) without the need for bursts of SF. A redshift-dependent SF timescale may affect our results. However, we note that our N-body simulation does not show any correlation between satellite mass and entry time, suggesting that environmental strangulation commences as frequently at later times when the redshift-dependent SF timescale is not short enough to exhaust the gas reservoir without the occurrence of SF bursts. 

A final caveat is that our argument for bursting and quenching is based on stellar metallicities measured with the {\sc firefly} spectral fitting code \citep{Comparat2017, Goddard2017, wilkinson_etal17}.
{\sc firefly} fits galactic spectra by using the stellar population synthesis model by \citet{maraston_stromback11} with the Medium-resolution Isaac Newton Telescope Library of Empirical Spectra \citep[MILES]{sanchezblazquez_etal06}.
The number of age and metallicity bins in the stellar population synthesis model and the spectral resolution of the MILES library affect the accuracy of the metallicity measurements.
Hence, it is important to keep an open mind to the possibility that $\Delta\log Z_*$ may decrease when a better sampling of the age-metallicity plane and higher-resolution spectra become available.
Changes so significant that they alter the qualitative picture presented in this article are unlikely but it is important to keep an open mind to the possibility that SFR increases before quenching may be less significant than what we measure in {\sc GalICS~2.0}.

\section{Conclusion}

We have used the {\sc GalICS~2.0} SAM to study how strangulation (the shutdown of gas accretion onto satellite galaxies) affects their chemical evolution. A key strength of our approach with respect to previous studies is that the ratio of the outflow rate to the SFR is not a free parameter. Mass-loading factors are determined by the requirement to reproduce the stellar mass function of galaxies.

In {\sc GalICS~2.0}, strangulation arises naturally because gas accretion is proportional to the growth of the DM halo and most haloes stop growing when they become subhaloes (often even before they are identified as such by the halo finder). As soon as the metallicity of the gas within a satellite is no longer diluted by the accretion of pristine gas from the intergalactic medium, its value begins to increase and so does the metallicity of the stars that form. Still, star formation must be significant for the stellar metallicity to increase substantially.

We find that, unless feedback is heavily suppressed (in which case {\sc GalICS~2.0} no longer reproduces the observed stellar mass function of galaxies and conditional stellar mass function of satellites), strangulation alone is not sufficient to explain the observed difference in metallicity between star-forming and passive galaxies (T18) because there is not enough star formation after the onset of strangulation. Moreover, the predicted fraction of passive galaxies ($\sim 20\%$) remains well below the observational range found by \citet[55 -- 95 percent]{wetzel_etal13}. Stripping the metal-poor gas in the outskirts of satellites does not solve the problem because it increases stellar metallicities by $0.02\,$dex at most.

Allowing small bursts of star formation at pericectric passages increases the passive fraction to 60 -- 70 percent. The bursts exhaust the satellites' gas reservoir and rapidly drive their metallicity to the high values observed. Numerous studies involving hydrodynamic simulations and observations have shown evidence for SFR increases in satellite galaxies \citep{Butcher1984, Bekki2003, Finn2005, Kronberger2008, Kapferer2009,Bekki2014, Paccagnella2017}. Physically their likely origin is compression by ram pressure which promotes the formation of molecular clouds \citep{Henderson2016} and triggers their collapse \citep{Bekki2003}. In our model, the bursts have a numerical origin. The halo finder overtruncates  subhaloes at pericentres because it assumes instantaneous tides (T17) and therefore underestimates their specific angular momenta. Artificially low disc sizes produce artificially enhanced SFRs. However, out artificial bursts mimic the behaviour that is expected physically.
They occur close to the pericentre and their strength grows with $M_{\rm h}/M_*$, where $M_{\rm h}$ is the virial mass of the host and $M_*$ is the stellar mass of the satellite, as expected from hydrodynamic simulations \citep{Bekki2014}. Hence, the median delay time between the onset of strangulation and quenching decreases with $M_{\rm h}$ and increases with $M_*$ at least up to $M_*\approx 10^{10}\,M_{\odot}$ in agreement with previous studies \citep{De-Lucia2012, Wetzel2015, Fillingham2015}.

\section*{Acknowledgements}

We thank James Trussler and Roberto Maiolino for providing the observational data and for their constructive comments. We also thank the anonymous referee for  several suggestions that improved the quality of this work.

\bibliographystyle{mn2e}

\bibliography{ref_av}
\end{document}